\documentclass[prapplied,twocolumn,superscriptaddress,amsmath,amssymb]{revtex4-2}
\usepackage{amsmath}
\usepackage{amsfonts}
\usepackage{amssymb}
\usepackage{graphicx}
\usepackage{mathrsfs}
\usepackage{color}
\usepackage{flushend}
\usepackage{txfonts}
\usepackage[titletoc]{appendix}
\usepackage[dvipdfm,colorlinks={true}]{hyperref}
\hypersetup{citecolor={blue}, filecolor={blue}, linkcolor={blue}, urlcolor={blue}}

\begin{document}
\title{Quantum Speed Limit under Brachistochrone Evolution}
\author{Fu-Quan Dou}
\email{doufq@nwnu.edu.cn}
\affiliation{College of Physics and Electronic Engineering, Northwest Normal University, Lanzhou, 730070, China}
\author{Min-Peng Han}
\affiliation{College of Physics and Electronic Engineering, Northwest Normal University, Lanzhou, 730070, China}
\author{Chuan-Cun Shu}
\email{cc.hu@csu.edu.cn}
\affiliation{Hunan Key Laboratory of Nanophotonics and Devices, Hunan Key Laboratory of Super-Microstructure and Ultrafast Process, School of Physics, Central South University,
Changsha, 410083, China}
\begin{abstract}
According to the Heisenberg uncertainty principle between time and energy fluctuation, a concept of the quantum speed limit (QSL) has been established to determine the minimum evolutionary time between quantum states. Considerable theoretical and experimental efforts are invested in obtaining the QSL time bounds in various scenarios. However, it remains a long-standing goal to derive a meaningful QSL bound for a general quantum problem. Here, we propose a geometrical approach to derive a QSL bound for closed and open quantum systems. By solving a quantum brachistochrone problem in the framework of the Riemannian metric, we show that the QSL between a given initial state to a final state is determined not only by the entire dynamics of the system but also by the individual dynamics of a critical parameter. We exemplify the utility of the new bound in three representative scenarios, demonstrating a pronounced advantage in finding a tight and meaningful QSL bound of a general quantum evolution problem.
\end{abstract}
\maketitle
\section{Introduction}
Besides its fundamental importance in quantum mechanics, quantum speed limit (QSL) that sets the maximal speed of evolution of a quantum system ultimately into a given target state is crucial in quantum computation \cite{lloyd2000ultimate}, quantum battery \cite{PhysRevLett.118.150601}, quantum control \cite{PhysRevLett.103.240501,PhysRevLett.118.100601}, quantum metrology \cite{PhysRevLett.119.010403,PhysRevLett.105.180402}, and other quantum technologies \cite{Deffner_2017, WOS:000383587100001}. For closed quantum systems governed by unitary time evolution with a time-independent Hamiltonian $\hat{H}$,  a seminal work by Mandelstam and Tamm (MT) first derived the QSL time in 1945  \cite{mandelstam1945uncertainty}, showing that the variance of the energy $\Delta E$ with respect to the initial state bounds the minimal evolution time by an inequality $\tau\ge\tau_{MT}=\pi\hbar/(2\Delta E) $ .
Margolus and Levitin (ML) provided the maximum speed of such unitary evolution with a different inequality $\tau\ge\tau_{ML}=\pi\hbar/(2E)$  in terms of the mean energy 
relative to the ground state energy \cite{MARGOLUS1998188}.\\ \indent
By combining the MT- and ML- type bounds, the QSL time between two orthogonal states of closed systems can be made tighter as $\tau\ge \mathrm{max}\{\tau_{MT}, \tau_{ML}\}$ \cite{Deffner_2017, WOS:000383587100001}.
Benefiting from various metrics, such as Fubini-Study metric \cite{PhysRevLett.65.1697}, trace distance \cite{PhysRevA.95.052104}, Bures distance \cite{PhysRevLett.110.050402,PhysRevLett.111.010402,PhysRevLett.126.180603}, and relative purity \cite{PhysRevLett.110.050403,Meng2015-sc}, considerable theoretical works have generalized the MT- and ML-type bounds originally  derived for the unitary evolution of pure states to different scenarios. For examples, this  includes the case of the mixed states \cite{MONDAL2016689,PhysRevLett.120.060409}, time-dependent Hamiltonian \cite{PhysRevLett.115.210402,PhysRevLett.129.140403,PhysRevLett.123.180403}, complex many-body system \cite{PhysRevResearch.2.023125,PhysRevResearch.2.032016,PhysRevX.9.011034}, open quantum systems \cite{PhysRevLett.127.100404,2022arXiv220702438W, PhysRevX.6.021031,PhysRevX.12.011038}, even for classical dynamics \cite{PRXQuantum.2.040349,PhysRevLett.120.070402, PhysRevLett.120.070401} and topological information about the system dynamics \cite{PhysRevLett.130.010402}. Due to the diversity and complexity of quantum problems of interest, establishing a unified QSL bound that is tight and attainable for both unitary and nonunitary evolutions remains  a long-standing goal.\\ \indent
In this work, we solve a general quantum brachistochrone (QB) problem in the Riemannian metric to derive a lower bound of the QSL time determined by the entire speed of the system and the individual speed of a critical parameter. The QB problem is a quantum analogy to Bernoulli's classical brachistochrone problem essentially concerns the determination of the fastest evolution path required to join a given initial state to a final state \cite{PhysRevLett.96.060503,PhysRevLett.114.170501,WOS:000880809900002}. Recently, investigating the QB problem with geometrical tools has generated considerable interest in searching for minimal evolution time in different quantum systems \cite{PhysRevX.11.011035,Kuzmak_2013,PhysRevLett.99.130502,PhysRevLett.101.230404,RevModPhys.90.015002,PhysRevLett.103.080502,PhysRevLett.103.080502,PhysRevA.100.062328,PhysRevA.103.012206}. Here, we present a theoretical framework to describe the QB evolution between two given states under general geometric representation through the Fubini-Study metric. To this end, we parameterize the metric and a family of quantum states with their corresponding amplitudes and phases for a quantum system. We ultimately obtain a lower bound of the QSL time in terms of the maximal entire speed of the system and the maximal individual speed of a critical parameter.
\\ \indent For illustrations, we first explore this new QSL bound in a closed quantum system with a nonlinear Landau-Zener (LZ) type Hamiltonian, commonly used for describing  a Bose-Einstein condensate (BEC) in a time-dependent two-level system under the mean-field approximation. We show how a QB evolution in geometrical parameter space (i.e., on the Bloch sphere) leads to the QSL time, explicitly providing a time-optimal control protocol consistent with that previously given by quantum optimal control theory (QOCT) simulations \cite{PhysRevLett.111.260501}, and experiments \cite{WOS:000300403700020}. As the second unitary evolution example, we examine a fast coherent transport of a trapped atom between two states with a significant spatial separation $d$. We find that the QSL time determined by the entire speed of the system is proportional to $\sqrt{d}$, in good agreement with a recent experiment and theoretical analysis considering the classical analog of the atom transport problem \cite{PhysRevX.11.011035}. We then generalize our approach to open quantum systems described by reduced density operators and apply it to the damped Jaynes-Cummings (JC) model, showing how the individual evolution speed of a critical parameter determines the QSL time. \\ \indent
\section{Quantum speed limit for closed systems}
To explain our method, we start to consider a closed quantum system that consists of $N$ eigenstates under
unitary evolution in the Hilbert space. The corresponding time-dependent wave-function of the system reads $|\psi(t)\rangle=\sum_{j=1}^Np_j(t)e^{i\varphi_j(t)}|j\rangle$ with the time-dependent amplitude $p_j(t)$ and phase $\varphi_j(t)$. Our QB problem refers to finding the time-optimal path connecting the initial state $|\psi(t=0)\rangle=|\psi_{0}\rangle$  and the target state $|\psi(t=\tau)\rangle=|\psi_{\tau}\rangle$ in the least possible time $\tau$, which corresponds to minimize the functional $\tau[\bullet]=\int d\mathcal{L}/\mathcal{V}$ with the infinitesimal distance $d\mathcal{L}$ and the quantum speed $\mathcal{V}$. To measure how fast the quantum system has evolved by the unitary operator, we define the Bures angle $\mathcal{L}(|\psi_0\rangle, |\psi(t)\rangle)=\arccos(|\langle\psi_0|\psi(t)\rangle|)$ to record the distance over the initial state $|\psi_0\rangle$ to $|\psi(t)\rangle$. The square of the infinitesimal distance [$d\mathcal{L}=\arccos(|\langle\psi(t)|\psi(t)+d\psi(t)\rangle|)$] can be written as the Riemannian metric of Fubini and Study \cite{PhysRevLett.65.1697,PhysRevD.23.357,PhysRevLett.72.3439,bengtsson2017,bengtsson2017}
\begin{equation}
d\mathcal{L}^{2}=\sum_{j=1}^{N}dp^{2}_{j}+\left[\sum_{j=1}^{N}p^{2}_{j}d\varphi^{2}_{j}-\left(\sum_{j=1}^{N}p^{2}_{j}d\varphi_{j}\right)^{2}\right].\label{Fubini1}
\end{equation}
Accordingly, the geometric quantum speed $\mathcal{V}=d\mathcal{L}/dt$ \cite{Deffner_2017} and the QB is determined by the parameters $p_{j}$ and $\varphi_{j}$. To obtain $\tau[\bullet]$ under arbitrary parameter description, we introduce a set of parameters $\{\lambda_{i}(t)\} (i=1,\ldots,r\leq2N)$, the infinitesimal distance can be further rewritten as $d\mathcal{L}^{2}=\sum_{\mu,\nu=1}^{r}g_{\mu\nu}d\lambda_{\mu}d\lambda_{\nu}$, where the components of the metric tensor read $g_{\mu\nu}=\sum_{i,j=1}^{N}[\delta_{ij}(\partial_{\mu} p_{i})(\partial_{\nu}p_{j})+(\delta_{ij}p_{j}^{2}-p_{i}^{2}p_{j}^{2})(\partial_{\mu}\varphi_{i})(\partial_{\nu}\varphi_{j})]$ with $\partial_{\mu,\nu}=\partial/\partial\lambda_{\mu,\nu}$.
\\ \indent
The functional $\tau[\bullet]$ then can be explicitly parameterized as
\begin{equation}
\tau[\bullet]=\int^{\zeta_{\tau}}_{\zeta_{0}}\frac{\sqrt{\sum\limits_{\mu,\nu=1}\limits^{r}g_{\mu \nu}\frac{d\lambda_{\mu}}{d\zeta}\frac{d\lambda_{\nu}}{d\zeta}}}{\partial_{t}\mathcal{L}(\lambda_{1},\ldots,\lambda_{r})}d\zeta=\int^{\zeta_{\tau}}_{\zeta_{0}}Td\zeta, \label{timefunctional}
\end{equation}
which satisfies the Euler-Lagrange (EL)  equation $d\partial_{\dot{\lambda_{i}}}T/d\zeta=\partial_{\lambda_{i}}T, i=1,\ldots,r$ with the natural parameter $\zeta$, often referred to as the QB equation \cite{PhysRevLett.96.060503,PhysRevLett.114.170501}. Thus, the QB problem turns to speed up the system evolution from the initial state $|\psi_{0}\rangle=|\psi\left(\lambda_{1}(0),\ldots,\lambda_{r}(0)\right)\rangle$ to the target state $|\psi_{\tau}\rangle=|\psi(\lambda_{1}(\tau),\ldots,\lambda_{r}(\tau))\rangle$ along the path $\mathcal{L}(|\psi_{0}\rangle,|\psi_{\tau}\rangle)=\int\partial_{\zeta}\mathcal{L}d\zeta$ on the Riemannian manifold $\mathcal{M}$. The quantum speed can be  parameterized as $\mathcal{V}(\lambda_{1},\ldots,\lambda_{r})\equiv{\partial_{t}\mathcal{L}(\lambda_{1},\ldots,\lambda_{r})}=\sqrt{\sum_{\mu,\nu=1}^{r}g_{\mu \nu}V_{\mu}V_{\nu}}$, where $V_{\mu,\nu}=d\lambda_{\mu,\nu}/dt$ denotes the evolution speed of a particular  parameter $\lambda_{\mu,\nu}$. For convenience, we call the two speeds $\mathcal{V}$ and $V_{\mu}$ as the global speed and the local speed, i.e., the entire speed of the system and the individual speed of a parameter, and define their maximum as $\mathcal{V}_{\max}$ and $|V_{\mu}|_{\max}$, respectively. As a result, the functional $\tau[\bullet]$ in terms of the maximal global speed $\mathcal{V}_{\max}$ can be given by
\begin{eqnarray}\label{QSL2}
  \tau[\bullet]\geq\frac{\mathcal{L}(|\psi_{0}\rangle,|\psi_{\tau}\rangle)}{\mathcal{V}_{\max}}.
\end{eqnarray}
To see how the functional $\tau[\bullet]$ depends on the local speed, we track the evolution of the parameter $\{\lambda_{i}(t)\}$ from $\{\lambda_{i}(0)\}$ to $\{\lambda_{i}(\tau)\}$ by selecting a natural parameter $\zeta=\lambda_{i}$, which is determined by the amplitudes $p_j$ and phases $\varphi_j$ of the quantum system, and therefore Eq. (\ref{timefunctional}) becomes
$\tau[\bullet]=\int d\lambda_{i}/V_{i}$. We then define the local geodesic $\mathfrak{L}(\lambda_{i}(0),\lambda_{i}(\tau))=|\lambda_{i}(\tau)-\lambda_{i}(0)|$ to describe the shortest distance between  $\lambda_{i}(0)$ and $\lambda_{i}(\tau)$. The maximum local speed $|V_{i}|_{\max}$ generates the minimal evolution time $\mathfrak{L}(\lambda_{i}(0),\lambda_{i}(\tau))/|V_{i}|_{\max}$ for a given parameter $\lambda_{i}$. For all parameters $\{\lambda_{i}(t)\}$,  the functional $\tau[\bullet]$ in terms of the maximal local speed can be obtained
\begin{equation} \label{QSL1}
\tau[\bullet]\geq\max\limits_{i=1,\dots,r}\left\{\frac{\mathfrak{L}(\lambda_{i}(0),\lambda_{i}(\tau))}{|V_{i}|_{\max}}\right\}=\frac{\mathfrak{L}(\lambda(0),\lambda(\tau))}{V_{\max}},
\end{equation}
which implies that we need to calculate the bounds of all parameters and find the maximum for a critical parameter to determine the local geodesic $\mathfrak{L}(\lambda(0),\lambda(\tau))$ and the maximal local speed  $V_\mathrm{max}$ in Eq. (\ref{QSL1}). Based these considerations, the QSL time takes the upper bound of Eqs. (\ref{QSL2}) and (\ref{QSL1})
\begin{eqnarray} \label{QSL}
\tau_{QSL}=\max\left\{\frac{\mathcal{L}(|\psi_{0}\rangle,|\psi_{\tau}\rangle)}{\mathcal{V}_{\max}},\frac{\mathfrak{L}(\lambda(0),\lambda(\tau))}{V_{\max}}\right\}.
\end{eqnarray}\\ \indent
Note that the two QSL time bounds defined by global and local speeds are independent and have different physical meanings for determining the minimal evolution time between the initial and final states. It implies that the QSL under QB evolution path depends on the entire dynamics of the system (i.e., the global speed) and the individual speed of a critical parameter (i.e., the local speed).  For practical applications, we must calculate the minimal times for each parameter and find the maximal one, which is further compared with the global one to determine a tight bound in Eq. (\ref {QSL}) as the QSL time. We should also note that the previous geometric approaches usually considered the geometric quantum speed limit determined by the global speed \cite{Deffner_2017}. Using QOCT and considering a QB problem in unitary evolution, the optimal Hamiltonian, as demonstrated in Ref. \cite{PhysRevLett.114.170501}, can be obtained by optimizing individual parameters, showing that the time-optimal control problem can be transformed into finding geodesic paths. However, the QOCT simulations usually cannot give a unified QSL bound as it depends on optimization algorithms and initial guesses. Our method considers both the maximal evolution speeds of the system and a critical parameter, showing that QSL could be determined when the system evolves along the geodesic at the maximal global speed, consistent with the finding in Ref. \cite{PhysRevLett.114.170501}.
\begin{figure}[htbp]
\centering
\includegraphics[width=0.32\textwidth]{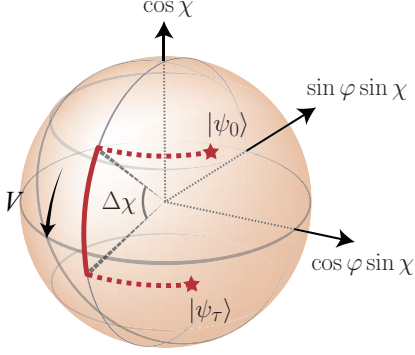}
\caption{The QB of the nonlinear Landau-Zener type system on the Bloch sphere. The upper star denotes the initial state and the lower star denotes the target state. The red curve denotes the evolution path of the state with the optimal protocol $\Gamma$. Here, the local geodesic $\mathfrak{L}(\chi_{0},\chi_{\tau})=\Delta\chi=|\chi_{\tau}-\chi_{0}|$ and the local speed $V$ describes the rate of $\chi$, i.e., the coupling strength $v$.}
\label{Fig1}
\end{figure}

\emph{Example 1: Nonliear Landau-Zener-type model.}--We now apply our method to a nonlinear LZ-type model, usually used to describe a Bose-Einstein condensate (BEC) in a time-dependent two-level system within the mean-field approximation \cite{PhysRevA.61.023402,PhysRevA.66.023404,PhysRevLett.91.230406,PhysRevA.89.012123,PhysRevA.93.043419,PhysRevA.98.022102,PhysRevA.94.023624}. The time evolution of the system is governed  by the Schr\"odinger equation with the Hamiltonian
\begin{equation}
\hat{H}(t)=\frac{[\Gamma(t)+c(|\psi_{2}|^{2}-|\psi_{1}|^{2})]}{2}\sigma_{z}+\frac{v}{2}\sigma_{x},
\end{equation}
where $\psi_{1}$ and $\psi_{2}$ are the probability amplitudes, the total
probability is conserved, i.e.,  $|\psi_{1}|^2+|\psi_{2}|^2=1$, $c$ describes the interaction between atoms, $v$ and $\Gamma(t)$ denotes the coupling strength and the energy bias between two levels controlled by the external fields. Our main task is to find the optimal protocol $\Gamma(t)$ and the QB to achieve the QSL. For the purpose of geometrical representation, we use the Bloch sphere to parameterize the wave function of the two-level system as $|\psi(t)\rangle=\cos(\chi(t)/2)e^{-i\varphi(t)/2}|1\rangle +\sin(\chi(t)/2)e^{i\varphi(t)/2}|2\rangle$, with a polar angle $\chi(t)\in [0,\pi]$ and an azimuth angle $\varphi(t)\in [0, 2\pi]$ \cite{Berry_2009}.
The evolution time can be expressed by the following form (details see Appendix \ref{appendixA})
\begin{equation}
d\tau =\frac{d\eta}{v\sqrt{1-\eta^{2}-\frac{[c(1-\eta^{2})-2\int\Gamma d\eta+C_{0}]^{2}}{4v^{2}}}},\label{expression}
\end{equation}
where $\eta=\cos\chi=|\psi_1|^2-|\psi_2|^2$ and $C_0$ is the integration constant determined by the initial values of $\varphi_0=\varphi(0)$ and $\chi_0=\chi(0)$. \\ \indent
We now select $\chi$ as the natural parameter that evolves from the initial angle $\chi(0)=\chi_0$ to the final angle $\chi(\tau)=\chi_\tau$ and then the functional $\tau[\bullet]$ can be written as
\begin{equation}
\tau[\bullet]=\int^{\chi_{\tau}}_{\chi_{0}}\frac{\sqrt{1+\dot\varphi^{2}\sin^{2}\chi}}{\frac{d}{dt}\mathcal{L}(\varphi,\chi)}d\chi=\int^{\chi_{\tau}}_{\chi_{0}}Td\chi.
\label{NTLtimefunctional}
\end{equation}
which satisfies the QB equation
$d\partial_{\dot{\varphi}}T/d\chi=\partial_{\varphi}T$. Solving the QB equation by the variational method, we can obtain the QB with $\varphi\equiv\pm\pi/2$.
Furthermore, the optimal protocol $\Gamma$ can be derived (see Appendix \ref{appendixA}),
\begin{equation}
\Gamma=c\cos\chi-\delta(\chi-\chi_{0})+\delta(\chi-\chi_{\tau}), \label{procotol}
\end{equation}
where $\delta(\chi)$ is the Dirac delta  function. This result in Eq. (\ref{procotol}) is consistent with a recent experimental result \cite{WOS:000300403700020} and  optimal control theory simulations \cite{PhysRevLett.111.260501}. The QSL occurs only if $\Gamma=c\cos\chi$ and $C_{0}=0$ with
\begin{eqnarray}
\tau_{QSL}=\frac{|\chi_{\tau}-\chi_{0}|}{v}, \label{QSLofNLZ}
\end{eqnarray}
 for both repulsive and attractive interactions. Of course, the QSL bound can be also obtained directly by Eq. (\ref{QSL}) (also see Appendix \ref{appendixA}).  Without loss of generality, Fig. \ref{Fig1} shows the QB for arbitrary initial and target states by considering the case $\chi_{0}<\chi_{\tau}$. We can see that the maximal local speed, related to the coupling strength $v$, determines the QSL. It implies that our new method solves a challenging task of finding QSL bounds of the nonlinear time-dependent systems. 

\begin{figure*}[t]
\centering
\includegraphics[width=0.76\textwidth]{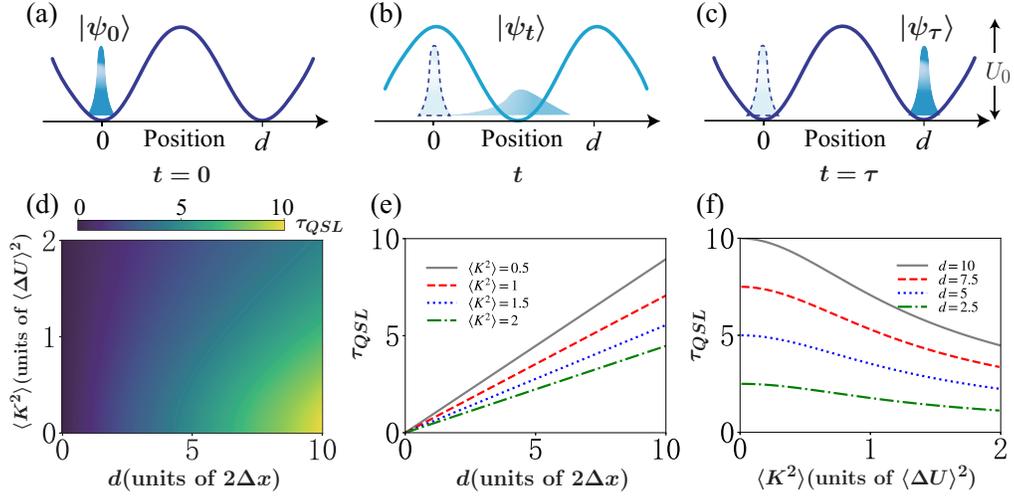}
\caption{The fast transport of a trapped atomic wave packet. (a)-(c) The initial state $|\psi_{0}\rangle$ in the left well of the potential is transferred over a distance $d$ by the time-dependent controls to the target state $|\psi_{\tau}\rangle$ in the right well of the potential. (d) The QSL time $\tau_{QSL}$ versus the distance $d$ and the mean square kinetic energy $\langle K^2\rangle$, (e) $\tau_{QSL}$ versus  $d$ for different values of $\langle K^2\rangle$, and (f) $\tau_{QSL}$ versus $\langle K^2\rangle$ for different values of $d$.}
\label{Fig2}
\end{figure*}

\emph{Example 2: Fast atomic transport problem.}--We now examine a problem by transporting  an atomic wave packet over a distance of 15 times its size, as demonstrated in a recent experiment  \cite{PhysRevX.11.011035}. A double-well potential is used to describe the transport of an atomic wave packet from the ground state
of the left well to the ground state of the right well, as shown in Figs. \ref{Fig2} (a)-(c). The time evolution of the system is governed by the Hamiltonian $\hat{H}(t)=\hat{p}^{2}/2m+\hat{U}(x,t)$ with the kinetic energy operator $\hat{p}^{2}/2m$ and the potential energy operator $\hat{U}(x,t)$ including the time-dependent interaction with control fields. Under the coordinate representation, the wave function of the $j^{th}$ state can be written as $\psi_{j}(x,t)=p_{j}(x, t)e^{i\varphi_{j}(x, t)}$. Using the harmonic approximation, we have $\partial_{t}p_{j}\approx-\left(\partial^{2}\varphi/\partial x^{2}\right)p_{j}/2m$, and $\partial_{t}\varphi_{j}\approx U(j\delta x,t)$ (the details is shown in the Appendix \ref{appendixB}).
According to the the position-momentum uncertainty relation $\Delta p\geq1/2\Delta x$ and Eq. (\ref{QSL}), the QSL can be specified by the global bound
\begin{equation}
\tau_{QSL}= \frac{d}{2\Delta x}\frac{1}{\sqrt{\langle K^{2}\rangle+(\Delta U)^{2} }_{\max}},\label{QSLexample2}
\end{equation}
where $d$ is the transport distance, $\langle K^2\rangle$ is the mean square kinetic energy, $\Delta U=\sqrt{\langle U^{2}\rangle-\langle U\rangle^{2}}$ and $d/2\Delta x$ denotes the Fubini-Study geodesic.\\ \indent
Figures \ref{Fig2} (d)-(f) show the dependence of the QSL on the distance $d$ and the mean square kinetic energy $\langle K^2\rangle$. Different from the conclusion given by MT bound that QSL time has nothing to do with transport distance, Figs.
 \ref{Fig2} shows that the longer the transport distance corresponds to the longer the QSL time. The QSL time $\tau_{QSL} $ is proportional to the Fubini-Study geodesic and is inversely proportional to the mean square kinetic energy, for which the bound in terms of the global speed can be understood as the constraint on the system.
By analysing the remote transmission and fixing trapped depth, we can further find that the global estimation provides a tighter bound than the local estimation. Interestingly, by considering an experimentally used conveyor belt potential $\hat{U}(x,t)=U_{0}\cos^{2}\{2\pi[\hat{x}-x_{control}(t)]/\lambda\}$
with the optical lattice wavelength $\lambda$ and the location of the potential center $x_{control}(t)$, the QSL time can be derived analytically with (details in Appendix \ref{appendixB})
\begin{equation}
  \tau_{QSL}=\sqrt{\frac{m\lambda^{2}d}{4\pi^{2} U_{0}\Delta x}}\propto\sqrt{d},
\end{equation}
which is consistent with a theoretical analysis in Ref. \cite{PhysRevX.11.011035} by using the classical analog of the atom transport model.

\section{Quantum speed limit for open systems}
We now extend our method to open quantum systems. The Bures angle in terms of the density matrix operator can be written as $\mathcal{L}(\rho_{0},\rho_{\tau})=\arccos\Big(\mathrm{Tr}\sqrt{\sqrt{\rho_{0}}\rho_{\tau}\sqrt{\rho_{0}}}\Big)$, where the initial density matrix  $\rho(0)=\rho_{0}$ and the target density matrix $\rho(\tau)=\rho_{\tau}$ with $\rho(t)=\sum_{j}\tilde{p}_{j}(t)|j(t)\rangle\langle j(t)|$ \cite{UHLMANN1993253}. To derive the QSL under arbitrary parameter description, we parameterize the square of the infinitesimal distance $d\mathcal{L}^{2}=\sum_{\mu,\nu=1}^{r}g_{\mu\nu}d\lambda_{\mu}d\lambda_{\nu}$ as the tensor form, where the components of the metric tensor becomes \cite{PhysRevX.6.021031}
\begin{equation}
g_{\mu\nu}=\frac{1}{4}\sum_{j}\frac{\partial_{\mu}\tilde{p}_{j}\partial_{\nu}\tilde{p}_{j}}{\tilde{p}_{j}}-\sum_{j<k}\frac{(\tilde{p}_{j}-\tilde{p}_{k})^{2}\langle j|\partial_{\mu}|k\rangle\langle k|\partial_{\nu}|j\rangle}{\tilde{p}_{j}+\tilde{p}_{k}}.
\end{equation}
Following a similar analysis to the closed systems, our geometric framework can be applied to open systems with $\rho_{0}$ and $\rho_{\tau}$ and the corresponding QSL bound can be given by (see Appendix \ref{appendixC})
\begin{equation}\label{QSLopen}
\tau_{QSL}=\max\left\{\frac{\mathcal{L}(\rho_{0},\rho_{\tau})}{\mathcal{V}_{\max}},\frac{\mathfrak{L}(\mathcal{X}(0),\mathcal{X}(\tau))}{V_{\max}}\right\},
\end{equation}
where $\mathcal{X}$ is determined by the parameters ${\tilde{p}_j}$ and the eigenstate $|j\rangle$.\\ \indent
\begin{figure}[t]
\centering
\includegraphics[width=0.32\textwidth]{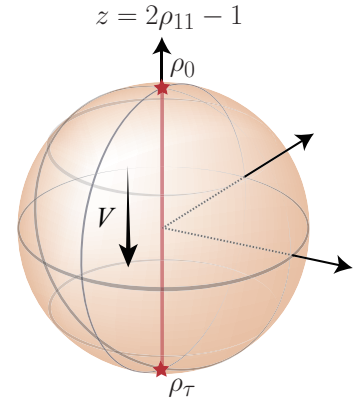}
\caption{ Evolution on the Bloch sphere. The red line denotes the evolution path and stars are the initial state and target state. The local geodesic $\mathfrak{L}(z(0),z(\tau))$ is the diameter along $z$-axis and the local speed $V=\partial_{t}z$ is the rate of $z=2\rho_{11}-1$.}
\label{fig3}
\end{figure}
\emph{Example 3: Damped Jaynes-Cummings model.}--To examine Eq. (\ref{QSLopen}) in open quantum systems, we take the damped JC model as an example, which is usually is used to describe a single two-level atom interacting with a single quantized field mode \cite{PhysRevA.59.1633,PhysRevA.55.2290}.
The nonunitary generator of the reduced dynamics of the system reads
\begin{equation}
L_{t}[\rho(t)]=\gamma(t)\left[\sigma_{-}\rho(t)\sigma_{+}-\frac{1}{2}\sigma_{+}\sigma_{-}\rho(t)-\frac{1}{2}\rho(t)\sigma_{+}\sigma_{-}\right],
\end{equation}
where
$\sigma_{\pm}=\sigma_{x}\pm i\sigma_{y}$ with  Pauli operators $\sigma_{x,y}$ and $\gamma(t)$ denotes  the time-dependent decay rate from $\rho_{0}=|1\rangle\langle1|$ to $\rho_{\tau}=|0\rangle\langle0|$. For the atom-cavity system, we can use an effective Lorentzian spectral to describe the environment effect by  $J(\omega)=\gamma_{0}\lambda_{0}/2\pi[(\omega-\omega_{0})^{2}+\lambda_{0}^{2}]$
with  the spectral width $\lambda_{0}$, the couple strength $\gamma_{0}$  and the transition frequency $\omega_{0}$ of the two-level system, and then the corresponding decay rate $\gamma(t)$ can be explicitly defined these parameters $\lambda_{0}$, $\gamma_{0}$ and $\omega_{0}$ \cite{PhysRevA.59.1633,PhysRevA.55.2290}. By selecting $\rho_{11}$ or $z=2\rho_{11}-1\in[-1,1]$ as the natural parameter, we can further derive the QSL time bound as (details in Appendix \ref{appendixC})
\begin{equation}
\tau_{QSL}=\max\left\{\pi\left(\frac{\sqrt{\rho_{11}}\sqrt{1-\rho_{11}}}{\partial_{t}\rho_{11}}\right)_{\min},\frac{1}{|\partial_{t}\rho_{11}|_{\max}}\right\},
\end{equation}
where $\rho_{11}=\exp{(-\int_{0}^{t}\gamma(t')dt')}$. For the case of the weak-coupling regime with $\gamma_{0}\ll\lambda_{0}$, we have $\tau_{QSL}=1/\gamma_{0}$ for Markovian dnyamics. As for the limit $\gamma_{0}\gg\lambda_{0}$, the dynamics is non-Markovian and the QSL becomes
\begin{equation}
\tau_{QSL}=\frac{1}{|\partial_{t}\rho_{11}|_{\max}}=\frac{1}{|\sigma_{t}|_{\max}}=\frac{2}{\sqrt{2\gamma_{0}\lambda_{0}-\lambda_{0}^2}},
\end{equation}
where $\sigma_{t}=\partial_{t}\rho_{11}$ 
describes the rate of the information backflow from the environment to the system \cite{PhysRevLett.103.210401} and the maximal local speed corresponds to its maximal value.
We can see an interesting phenomenon that the non-Markovian effects can
speed up quantum evolution as discussed in previous works \cite{PhysRevLett.111.010402}. Figure \ref{fig3} shows the evolution of the system on the Bloch sphere. We find that the QSL time is determined by the local speed and the non-Markovian effects increase the local speed and further speed up the information back-flow from the environment to the system, see details in Appendix \ref{appendixC}.  
\\ \indent
\section{Conclusions}
We presented a geometrical approach to obtain the minimum evolution time of a quantum system under quantum brachistochrone (QB) evolution between states. By solving a general QB problem in the framework of the Riemannian metric,  we showed that a critical parameter in the geometric representation also determines the QSL bounds. As a result, the lower QSL time bound along QB evolution depends on the entire speed of the system and the individual speed of a critical parameter. We examined this new method for unitary and nonunitary quantum processes in three scenarios, demonstrating that our new QSL bounds can give meaningful predictions in the context of realistic quantum control problems. \\ \indent
The unified QSL time for a given quantum system could be utilized as the cost functional for estimating the fastest evolution time between states in quantum computation and simulations. By setting the QSL time as the total evolution time of the system, constrained QOCT algorithms or optimal control conditions can be applied to searching for optimized control fields in time-optimal quantum control problems \cite{Glaser2015,PhysRevA.93.043410,PhysRevA.93.053418,PhysRevA.100.023409,PhysRevA.104.013108,PhysRevLett.130.043604}. The present method may open up new applications in quantum technology, including quantum control, quantum metrology and quantum information.

\section*{Acknowledgments}
The work is supported by the National Natural Science Foundation of China (NSFC) under Grant No. 12075193. C.-C.S. is supported in part by the NSFC under Grant  Nos.12274470 and 61973317 and the Natural Science Foundation of Hunan Province for Distinguished Young Scholars under Grant No. 2022JJ10070.

\appendix
\section{Nonliear Landau-Zener-type model}\label{appendixA}
\subsection{Dynamics equations under the Bloch parametrization}
The Hamiltonian of the nonlinear LZ-type two-level system reads $\hat{H}=p\sigma_{z}+q\sigma_{x}$
with $p=[\Gamma(t)+c(|\psi_{2}|^{2}-|\psi_{1}|^{2})]/2$ and $q=v/2$.
For $|\psi_{1}|^{2}$ we have $d|\psi_{1}|^{2}/dt=(d\psi_{1}^{\ast}/dt) \psi_{1}+\psi_{1}^{\ast}(d\psi_{1}/dt)$. Submitting it into the Schr\"odinger equation, we can obtain
\begin{equation}
\frac{d|\psi_{1}|^{2}}{dt}=i\psi_{1}(p\psi_{1}^{\ast}+q\psi_{2}^{\ast})-i\psi_{1}^{\ast}(p\psi_{1}+q\psi_{2})
=iq(\psi_{1}\psi_{2}^{\ast}-\psi_{1}^{\ast}\psi_{2}).\label{B2}
\end{equation}
Under the Bloch parametrization with $\psi_{1}=\cos(\chi/2)e^{i\varphi_{1}}$ and $\psi_{2}=\sin(\chi/2)e^{i\varphi_{2}}$, we have
\begin{equation}
\frac{d|\psi_{1}|^{2}}{dt}
=q\sin\chi\sin\varphi,\label{B3}
\end{equation}
with the population difference  $\cos\chi=|\psi_{1}|^{2}-|\psi_{2}|^{2}$ and the relative phase $\varphi=\varphi_{2}-\varphi_{1}$. Equation~(\ref{B3}) can be further written as
\begin{equation}
\frac{d|\psi_{1}|^{2}}{dt}=\frac{d\left[\cos^{2}\left(\frac{\chi}{2}\right)\right]}{dt}=-\frac{\sin\chi}{2}\frac{d\chi}{dt}.\label{B4}
\end{equation}
Combining with Eqs.~ (\ref{B3}) and (\ref{B4}), we obtain $d\chi/dt=-2q\sin\varphi.$ For $\psi_{1}$, we have
\begin{eqnarray}
\begin{split}
\frac{d\psi_{1}}{dt}&=-\frac{1}{2}\sin\left(\frac{\chi}{2}\right)e^{i\varphi_{1}}\frac{d\chi}{dt}+i\cos\left(\frac{\chi}{2}\right)e^{i\varphi_{1}}\frac{d\varphi_{1}}{dt},\\
\end{split}
\end{eqnarray}
which can be used to obtain
\begin{equation}
\frac{d\varphi_{1}}{dt}=-p-q\tan\left(\frac{\chi}{2}\right)e^{i\varphi}-\frac{i}{2}\tan\left(\frac{\chi}{2}\right)\frac{d\chi_{1}}{dt}.\label{B7}
\end{equation}
Similarly, for $\psi_{2}$ we have
\begin{equation}
\frac{d\psi_{2}}{dt}=\frac{1}{2}\cos\left(\frac{\chi}{2}\right)e^{i\varphi_{2}}\frac{d\chi}{dt}+i\sin\left(\frac{\chi}{2}\right)e^{i\varphi_{2}}\frac{d\varphi_{2}}{dt},
\end{equation}
where  $\varphi_{2}$ satisfies
\begin{equation}
\frac{d\varphi_{2}}{dt}=p-q\cot\left(\frac{\chi}{2}\right)e^{-i\varphi}+\frac{i}{2}\cot\left(\frac{\chi}{2}\right)\frac{d\chi}{dt}.\label{B9}
\end{equation}

As a result, we obtain
\begin{equation}
\frac{d\varphi}{dt}=\frac{d\varphi_{2}}{dt}-\frac{d\varphi_{1}}{dt}=2p-2q\cos\varphi\cot\chi.
\end{equation}
The dynamics equations become
\begin{equation}
\frac{d\chi}{dt}=-v\sin\varphi, \frac{d\varphi}{dt}=\Gamma-\cos\chi\left(c+\frac{v\cos\varphi}{\sin\chi}\right). \label{BDiffenentialEquations}
  \end{equation}

In order to derive the QB of the nonlinear LZ type two-level system, we calculate the analytical expressions of the time-dependent evolution of $\chi(t)$ and $\varphi(t)$.
Under the case $\Gamma=0$, we find that Eq. (\ref{BDiffenentialEquations}) satisfies $\sin\chi=-2v\cos\varphi/c$.

For $\Gamma\neq0$, we assume that
\begin{equation}
\sin\chi=-\frac{2v\cos\varphi}{c+f},\label{BEx2}
\end{equation}
with the undetermined function $f$. To determine $f$, we combine with Eqs.~(\ref{BEx2}) and (\ref{BDiffenentialEquations}) then we obtain $\dot f=-2f\tan\chi-2\Gamma/\sin\chi$, where the dot denotes the derivative with respect to $\chi$. We further gives
\begin{equation}
f=\frac{-2\int\Gamma \sin\chi d\chi+C_{0}}{\sin^{2}\chi},\label{f}
\end{equation}
where $ C_{0}$ is the integration constant determined by the initial phase $\varphi(0)$ and initial angle $\chi(0)$. For arbitrary $\Gamma$, $\chi(t)$ and $\varphi(t)$ satisfies
\begin{equation}
\sin\chi=-\frac{2v}{c+\frac{-2\int\Gamma\sin\chi d\chi+C_{0}}{\sin^{2}\chi}}\cos\varphi.\label{BExpression}
\end{equation}
By employing Eq. (\ref{BDiffenentialEquations}), the evolution time can be further calculated by the following form
\begin{equation}
d\tau =\frac{d\eta}{v\sqrt{1-\eta^{2}-\frac{[c(1-\eta^{2})+2\int\Gamma d\eta+C_{0}]^{2}}{4v^{2}}}},\label{Bexpression}
\end{equation}
where $\eta=\cos\chi$. 

\subsection{Derivation of the QB and optimal protocol}
To derive QB, it is necessary to simplify the geometric quantum speed, which corresponds to the evolution speed of the point on the Bloch sphere. To this end, we have
$\mathcal{V}^{2}=|d\mathcal{L}/dt|^{2}=\left(d\varphi/dt\right)^{2}+\left(d\chi/dt\right)^{2}\sin^{2}\chi$.
By consdiering Eq.~(\ref{BDiffenentialEquations}), $\mathcal{V}^{2}$ can be rewritten as
\begin{equation}
\mathcal{V}^{2}=v^{2}(1-\cos^{2}\varphi)+\left[\Gamma-\cos\chi\left(c+\frac{v\cos\varphi}{\sin\chi}\right)\right]^{2}\sin^{2}\chi.
\end{equation}
Using Eq.~(\ref{BEx2}), we obtain
\begin{equation}
\mathcal{V}^{2}=\!v^{2}-\frac{(c+f)^{2}\sin^{2}\chi}{4}+\left(\Gamma-\frac{c\cos\chi}{2}+\frac{f\cos\chi}{2}\right)^{2}\sin^{2}\chi.\label{BV2}
\end{equation}
The EL equations become
$d\partial_{\dot{\varphi}}T/d\chi=\partial_{\varphi}T$,
where
\begin{equation}
\begin{split}
\frac{\partial T}{\partial\dot\varphi}&=\frac{\dot{\varphi}\sin^{2}\chi}{\dot{\mathcal{L}}\mathcal{V}},\;\frac{\partial T}{\partial\varphi}=0. \label{BQBE}
\end{split}
\end{equation}
The solution of Eq.~(\ref{BQBE}), i.e., QB of the nonlinear LZ type system can be given by
\begin{equation}
\varphi=\pm\frac{\pi}{2},\;\dot\varphi=0. \label{BQB}
\end{equation}
Combining with Eq.~(\ref{BExpression}), the optimal protocol $\Gamma$ is obtained:
\begin{equation}
\Gamma=c\cos\chi-\delta(\chi-\chi_{0})+\delta(\chi_{\tau}-\chi). \label{Bprocotol}
\end{equation}

For the Delta pulse, the speed of $\varphi(t)$ is infinity, indicating that the maximal global speed  becomes infinity and the global bound gives
\begin{equation}
\tau\geq\frac{\mathcal{L}(|\psi_{0}\rangle,|\psi_{\tau}\rangle)}{\sqrt{(\frac{d\varphi}{dt})^{2}+(\frac{d\chi}{dt})^{2}\sin^{2}\chi}_{\max}}=0.\label{BGlobal1}
\end{equation}
If we select $\varphi$ as the natural parameter, the corresponding local bound can be derived:
\begin{equation}
\tau\geq\frac{|\varphi_{\tau}-\varphi_{0}|}{|\partial_{t}\varphi|_{\max}}=0. \label{BQSLofNLZ}
\end{equation}
If we select $\chi$ as the natural parameter, according to the local bound and Eq. (\ref{BDiffenentialEquations}), we
obtain
\begin{equation}
\tau\geq\frac{|\chi_{\tau}-\chi_{0}|}{v}.
\end{equation}
It implies that the evolution speed of the parameter $\chi$ determines the QSL time for the nonlinear Landau-Zener-type model.

\section{Fast atomic transport}\label{appendixB}
Under the coordinate representation, $\psi_{j}=p_{j}e^{i\varphi_{j}}$ is introduced to denote the eigenstate of the $j^{th}$ location. These locations we considered are separated by the infinitely small distance $\delta x$. To derive the global speed, we calculate $\partial_{t}p_{j}$ and $\partial_{t}\varphi_{j}$.
Using the harmonic approximation, the kinetic energy term can be written as
\begin{eqnarray}
\begin{split}
\frac{\hat{p}^{2}}{2m}\psi_{j}&=-\frac{1}{2m}\frac{\partial^{2}}{\partial x^{2}}\psi_{j}\\
&\approx-\frac{1}{2m}\frac{\psi_{j+1}+\psi_{j-1}-2\psi_{j}}{\delta x^{2}}\\
&\approx-\frac{1}{2m}p_{j}\frac{e^{i\varphi_{j+1}}+e^{i\varphi_{j-1}}-2e^{i\varphi_{j}}}{\delta x^{2}}\\
&=-\frac{1}{2m}\psi_{j}\frac{e^{i(\varphi_{j+1}-\varphi_{j})}+e^{i(\varphi_{j-1}-\varphi_{j})}-2}{\delta x^{2}} \\ \label{Ckinetic}
&\approx-\frac{i}{2m}\left(\frac{\partial^{2}\varphi}{\partial x^{2}}\right)_{x=j\delta x}\psi_{j},
\end{split}
\end{eqnarray}
Based on Schr\"odinger equation
$i\partial_{t}\psi_{j}=(\hat{p}^{2}/2m+U)\psi_{j}$ and the relationship
$i\partial_{t}\psi_{j}=i[\partial_{t}p_{j}e^{i\varphi_{j}}+ip_{j}(\partial_{t}\varphi_{j})e^{i\varphi_{j}}]$, we can further obtain
\begin{equation}
\partial_{t}p_{j}\approx-\frac{1}{2m}\left(\frac{\partial^{2}\varphi}{\partial x^{2}}\right)p_{j}, \partial_{t}\varphi_{j}\approx U(j\delta x,t),\label{CODE2}
  \end{equation}
where $\partial_{t}p_{j}$ corresponds to the kinetic energy and $\partial_{t}\varphi_{j}$ is related to the potential energy. By inserting Eq. (\ref{CODE2}) into the Fubini-Study metric,  the global speed can be given by,
\begin{equation}
\partial_{t}\mathcal{L}=\sqrt{\sum_{i=1}^{N}\left(\frac{dp_{i}}{dt}\right)^{2}+\sum_{i=1}^{N}p^{2}_{i}\left(\frac{d\varphi_{i}}{dt}\right)^{2}-\left(\sum_{j=1}^{N}p^{2}_{j}\frac{d\varphi_{j}}{dt}\right)^{2}}.
\label{CFubini}
\end{equation}
Under the limit $\delta x\to0$, we can obtain
\begin{equation}
\partial_{t}\mathcal{L}=\sqrt{\langle K^{2}\rangle+(\Delta U)^{2}},\label{CV}
\end{equation}
where $\langle K^{2}\rangle$ is the mean value of the square of kinetic energy, $\Delta U=\sqrt{\langle U^{2}\rangle-\langle U\rangle^{2}}$ is the variance of the potential $U$.
To obtain the Fubini-Study geodesic, we select $x$ as the natural parameter. We can derive
\begin{equation}
\partial_{x}\mathcal{L}=\sqrt{\sum_{i=1}^{N}\left(\frac{dp_{i}}{dx}\right)^{2}+\sum_{i=1}^{N}p^{2}_{i}\left(\frac{d\varphi_{i}}{dx}\right)^{2}-\left(\sum_{j=1}^{N}p^{2}_{j}\frac{d\varphi_{j}}{dx}\right)^{2}},
\end{equation}
which corresponds to the variation of the Fubini-Study geodesic by moving the state in the parameter space from $x$ to $x+dx$.
The harmonic approximation yields $\partial_{x}p_{j}\propto \partial^{3}U/\partial x^{3}\approx0$, which gives $-i\partial_{x}\psi_{j}\!=\!-i(\partial_{x} p_{j})e^{i\varphi_{j}}+(\partial_{x}\varphi_{j})p_{j}e^{i\varphi_{j}}\!\approx\!(\partial_{x} \varphi_{j})\psi_{j}$.
Under the limit $\delta x\to0$, we have
\begin{equation}
\lim\limits_{N\to\infty}\sqrt{\sum_{j=1}^{N}p^{2}_{j}\left(\frac{d\varphi_{j}}{dx}\right)^{2}-\left(\sum_{j=1}^{N}p^{2}_{j}\frac{d\varphi_{j}}{dx}\right)^{2}}=\Delta p,\label{CL}
\end{equation}
where $\Delta p$ is the variance of momentum $p$.
With the position-momentum uncertainty relation $\Delta p\geq1/2\Delta x$, Eqs.~(\ref{CV}) and (\ref{CL}) give the QSL
\begin{equation}
\tau_{QSL}= \frac{d}{2\Delta x}\frac{1}{\sqrt{\langle K^{2}\rangle+(\Delta U)^{2} }_{\max}},\label{CQSLexample2}
\end{equation}
where $\mathcal{L}=d/2\Delta x$ is the Fubini-Study geodesic.
Note that the evolution path becomes the geodesic when the state of the system is the coherent state. For the potential $U(x,t)=U_{0}\cos^{2}\{2\pi[\hat{x}-x_{control}(t)]/\lambda\}$, the term of the kinetic energy can be explicitly calculated by using Eq. (\ref{CODE2})
\begin{eqnarray}
\begin{split}
\langle K^{2} \rangle&=\sum_{i=1}^{N}\left[\frac{1}{2m}\left(\frac{\partial^{2}\varphi}{\partial x^{2}}\right)\right]^{2}p_{j}^{2}\leq \left(\frac{k^{2}}{2m}U_{0}\tau\right)^{2},
\end{split}
\end{eqnarray}
where $k=2\pi/\lambda$. As a result, we have
\begin{equation}
\tau_{QSL}^{2}\geq\frac{d}{2\Delta x\sqrt{\left(k^{2}U_{0}/2m\right)^{2}+[(\Delta U)_{\max}/\tau_{QSL}]^{2}}}.\label{CexperQSL}
\end{equation}
For the remote transmission and fixed trapped depth, $[(\Delta U)_{\max}/\tau_{QSL}]^{2}\approx0$. Thus the QSL becomes
\begin{equation}
\tau_{QSL}=\sqrt{\frac{m\lambda^{2}}{4\pi^{2} U_{0}\Delta x}}\sqrt{d}\propto\sqrt{d}.\label{CQSLexp}
\end{equation}
For the experiment in Ref. \cite{PhysRevX.11.011035}, the harmonic oscillation frequency takes
\begin{equation}
\omega_{HO}=\frac{2\pi}{\tau_{HO}}=2\pi\sqrt{\frac{2U_{0}}{m\lambda^{2}}}\label{frequency},
\end{equation}
which gives the QSL
\begin{equation}
\tau_{QSL}=\sqrt{\frac{\lambda}{8\pi\Delta x}}\sqrt{\frac{2n}{\pi}\tau_{HO}}\approx1.17\sqrt{\frac{2n}{\pi}\tau_{HO}},
\end{equation}
where $n=2d/\lambda$ with $\lambda=866$ nm and $\Delta x\approx25$ nm in the experiment \cite{PhysRevX.11.011035}.

We then select $p_{j}$ as the natural parameter and the local QSL bound can be derived as
\begin{equation}
\tau\geq\frac{\mathfrak{L}(p_{j}(0),p_{j}(\tau))}{V_{\max}}=\frac{|p_{j}(0)-0|}{|\partial_{t}p_{j}|_{max}}=\frac{m\lambda^{2}}{2\pi^{2}U_{0}\tau}.
\end{equation}
We further obtain
\begin{equation}
\tau\geq\sqrt{\frac{m\lambda^{2}}{2\pi^{2} U_{0}}},
\end{equation}
which describes the minimal necessary time of the complete transfer of the $j^{th}$ eigenstate. For remote transmission, the global bound is coincident with the result given by the experiment and the QSL certainly increases with $d$. As a result, the $\tau_{QSL}$ is given by
\begin{equation}
\tau_{QSL}=\max\left\{\sqrt{\frac{m\lambda^{2}d}{4\pi^{2} U_{0}\Delta x}},\sqrt{\frac{m\lambda^{2}}{2\pi^{2} U_{0}}}\right\},\label{CGlobal2}
\end{equation}
which illustrates that the global bound is equal to the local bound multiplied by the Fubini-Study geodesic factor $\sqrt{d/2\Delta x}$.
\section{Extension to open quantum systems} \label{appendixC}
\subsection{Quantum speed limit time bounds}
For the open quantum system, the Bures angle in terms of the density matrix operator can be written as $\mathcal{L}(\rho_{0},\rho_{\tau})=\arccos(\mathrm{Tr}\sqrt{\sqrt{\rho_{0}}\rho_{\tau}\sqrt{\rho_{0}}})$, where the initial density matrix  $\rho(0)=\rho_{0}$ and the target density matrix $\rho(\tau)=\rho_{\tau}$ with $\rho(t)=\sum_{j}\tilde{p}_{j}(t)|j(t)\rangle\langle j(t)|$. To derive the QSL under arbitrary parameter description, we parameterize the square of the infinitesimal distance as the tensor form  \cite{PhysRevX.6.021031}
\begin{equation}
d\mathcal{L}^{2}=\sum\limits_{\mu,\nu}g_{\mu\nu}d\lambda_{\mu}d\lambda_{\nu},\label{Dopenmetric}
\end{equation}
where
\begin{equation}
g_{\mu\nu}=\frac{1}{4}\sum_{j}\frac{\partial_{\mu}\tilde{p}_{j}\partial_{\nu}\tilde{p}_{j}}{\tilde{p}_{j}}-\sum_{j<k}\frac{(\tilde{p}_{j}-\tilde{p}_{k})^{2}\langle j|\partial_{\mu}|k\rangle\langle k|\partial_{\nu}|j\rangle}{\tilde{p}_{j}+\tilde{p}_{k}}
\end{equation}
with $\partial_{\mu,\nu}=\partial/\partial\lambda_{\mu,\nu}$. Thus the functional $\tau[\bullet]$ then can be explicitly parameterized as
\begin{equation}
\tau[\bullet]=\int^{\zeta_{\tau}}_{\zeta_{0}}\frac{\sqrt{\sum\limits_{\mu,\nu=1}\limits^{r}g_{\mu \nu}\frac{d\lambda_{\mu}}{d\zeta}\frac{d\lambda_{\nu}}{d\zeta}}}{\partial_{t}\mathcal{L}(\lambda_{1},\ldots,\lambda_{r})}d\zeta.\label{Dtimefunctional}
\end{equation}
The QBE of the open quantum systems refers to the EL equations $d\partial_{\dot{\lambda_{i}}}T/d\zeta=\partial_{\lambda_{i}}T, i=1,\ldots,r$ with the natural parameter $\zeta$.
For the maximal global speed $\mathcal{V}_\mathrm{max}$ we have
\begin{eqnarray}
  \tau[\bullet]\geq\frac{\mathcal{L}(\rho_{0},\rho_{\tau})}{\mathcal{V}_{\max}}.\label{DopenQSL1}
\end{eqnarray}
Similar to the closed systems, we track the evolution of the parameter $\{\lambda_{i}(t)\}$ from $\{\lambda_{i}(0)\}$ to $\{\lambda_{i}(\tau)\}$ by selecting the natural parameter $\zeta=\lambda_{i}$ ($\lambda_{i}$ is determined by $\tilde{p}_{j}(t)$ and $|j(t)\rangle\langle j(t)|$) and therefore Eq. (\ref{Dtimefunctional}) becomes
$\tau[\bullet]=\int d\lambda_{i}/v_{i}$. For all parameters $\{\lambda_{i}(t)\}$, the functional $\tau[\bullet]$ in terms of the maximal local speed can be obtained
\begin{equation}
\tau[\bullet]\geq\max\limits_{i=1,\dots,r}\left\{\frac{\mathfrak{L}(\lambda_{i}(0),\lambda_{i}(\tau))}{|V_{i}|_{\max}}\right\}=\frac{\mathfrak{L}(\mathcal{X}(0),\mathcal{X}(\tau))}{V_\mathrm{max}},
\label{DopenQSL2}
\end{equation}
Thus the QSL time takes the upper bound of Eqs. (\ref{DopenQSL1}) and (\ref{DopenQSL2})
\begin{eqnarray}
\tau_{QSL}=\max\left\{\frac{\mathcal{L}(\rho_{0},\rho_{\tau})}{\mathcal{V}_{\max}},\frac{\mathfrak{L}(\mathcal{X}(0),\mathcal{X}(\tau))}{V_\mathrm{max}}\right\}.
\end{eqnarray}
\subsection{Application to the damped Jaynes-Cummings model}
For the damped Jaynes-Cummings model, the decay rate can be explicitly written as
\begin{equation}
\gamma(t)=\frac{2\gamma_{0}\lambda_{0}\sinh(Dt/2)}{D\cosh(Dt/2)+\lambda_{0}\sinh(Dt/2)},
\end{equation}
where $D=\sqrt{|\lambda_{0}^2-2\gamma_{0}\lambda_{0}|}$.
We select the initial state is the excited state, i.e., $\rho_{0}=\begin{pmatrix} 1 & 0\\ 0 & 0  \end{pmatrix}$,
and the corresponding target state is $\rho_{\tau}=\begin{pmatrix} 0 & 0\\ 0 & 1  \end{pmatrix}$.
In this case, the evolution of the density operator can be given by
\begin{equation}
\rho(t)=\begin{pmatrix} \rho_{11} & 0\\ 0 & 1-\rho_{11}  \end{pmatrix}, \label{Doperator}
\end{equation}
where $\rho_{11}=|e^{-\int_{0}^{t} \gamma(t') dt'}|$.
For this two-level system, we select time $t$ as the parameter and Eq. (\ref{Dopenmetric}) gives the global speed
\begin{equation}
\mathcal{V}=|\partial_{t}\mathcal{L}|=\frac{|\partial_{t}\rho_{11}|}{2\sqrt{\rho_{11}}\sqrt{1-\rho_{11}}},
\end{equation}
Since the length of the geodesic between the initial state and the target state is $\mathcal{L}=\pi/2$, the global speed bound gives
\begin{equation}
\frac{\mathcal{L}(\rho_{0},\rho_{\tau})}{\mathcal{V}_{\max}}=\pi\left(\frac{\sqrt{\rho_{11}}\sqrt{1-\rho_{11}}}{|\partial_{t}\rho_{11}|}\right)_{\min}.
\end{equation}
To calculate the local speed, we select $\rho_{11}$ as the natural parameter. The local speed bound can be written as
\begin{equation}
\frac{\mathfrak{L}(\rho_{11}(0),\rho_{11}(\tau))}{V_{\max}}=\frac{1}{|\partial_{t}\rho_{11}|_{\max}}.\label{Dlocal}
\end{equation}

From the perspective of the Bloch vector, the density operator becomes
$\rho(t)=({\rm I}+\vec{r}\cdot\vec{\sigma})/2$,
where $\rm I$ is the identity matrix, $\vec{r}=(x,y,z)$, $\vec{\sigma}=(\sigma_{x},\sigma_{y},\sigma_{z})$. Eq. (\ref{Doperator}) can be further expressed as
\begin{equation}
\rho(t)=\frac{1}{2}\begin{pmatrix} 1+z & x-iy\\ x+iy & 1-z  \end{pmatrix}=\frac{1}{2}\begin{pmatrix} 1+z & 0\\ 0 & 1-z  \end{pmatrix}.
\end{equation}
We select $z$ as the natural parameter. The maximal local speed $V=|\partial_{t}z|_{\max}$ and the geodesic $\mathfrak{L}=|z_{0}-z_{\tau}|=2$. The local speed bound gives:
\begin{equation}
\frac{\mathfrak{L}(\rho_{11}(0),\rho_{11}(\tau))}{V_{\max}}=\frac{2}{|\partial_{t}z|_{\max}}.\label{DQSL2}
\end{equation}
Since $z=2\rho_{11}-1$, Eq.~(\ref{DQSL2}) is the same as Eq.~(\ref{Dlocal}).

As a result, the minimal evolution time bound  can be given by
\begin{equation}
\tau_{QSL}=\max\left\{\pi\left(\frac{\sqrt{\rho_{11}}\sqrt{1-\rho_{11}}}{|\partial_{t}\rho_{11}|}\right)_{\min},\frac{1}{|\partial_{t}\rho_{11}|_{\max}}\right\}.\label{QSL3}
\end{equation}
Obviously, $\rho_{11}(0)=1$ at  $t=0$ and the global speed becomes infinity. The global bound gives
\begin{equation}
\pi\left(\frac{\sqrt{\rho_{11}}\sqrt{1-\rho_{11}}}{|\partial_{t}\rho_{11}|}\right)_{\min}=0.
\end{equation}
Thus the local bound is sharper. The dynamics is Markovian for the weak coupling $\gamma_{0}<\lambda_{0}/2$. Specially for the limit $\gamma_{0}\ll\lambda_{0}$, we have $\gamma(t)\approx\gamma_{0}$ \cite{PhysRevLett.111.010402}. The couple strength $\gamma_{0}$ determines the local speed and the QSL can be  given
\begin{equation}
\tau_{QSL}=\frac{1}{|\partial_{t}\rho_{11}|_{\max}}=\frac{1}{\gamma_{0}}.
\end{equation}
For non-Markovian dynamics with the strong coupling $\gamma_{0}>\lambda_{0}/2$,  the corresponding decay rate reads
\begin{equation}
\gamma(t)=\frac{2\gamma_{0}\lambda_{0}\tan(Dt/2)}{D+\lambda_{0}\tan(Dt/2)}.
\end{equation}
\begin{figure}[htbp]
\centering
\includegraphics[width=0.405\textwidth]{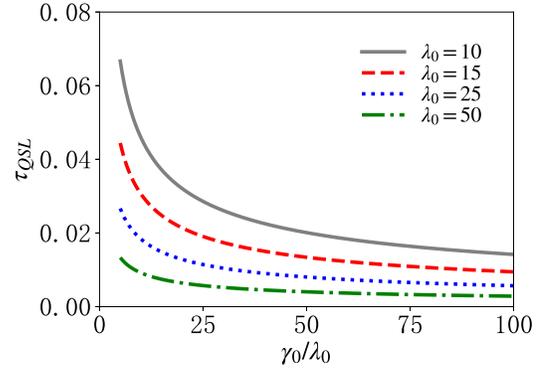}
\caption{QSL time as a function of $\gamma_{0}$ for different $\lambda_{0}$, which describes the acceleration effect from the non-Markovianity.}
\label{SMfig1}
\end{figure}
For $\gamma_{0}\gg\lambda_{0}$, we have $|\partial_{t}\rho_{11}|\approx D\sin(Dt)/2$ \cite{Meng2015-sc}. The corresponding local speed  becomes
$|\partial_{t}\rho_{11}|_{\max}=\sqrt{2\gamma_{0}\lambda_{0}-\lambda_{0}^2}/2$.
The QSL can be explicitly written as
\begin{equation}
\tau_{QSL}=\frac{1}{|\partial_{t}\rho_{11}|_{\max}}=\frac{2}{\sqrt{2\gamma_{0}\lambda_{0}-\lambda_{0}^2}}
=\frac{2}{\lambda_{0}\sqrt{2\frac{\gamma_{0}}{\lambda_{0}}-1}}\label{DQSLopen}.
\end{equation}

Figure  \ref{SMfig1} displays the QSL time as a function of $\gamma_{0}$ for different $\lambda_{0}$.
To describe acceleration effect, we introduce the non-Markovianity \cite{PhysRevLett.103.210401}
\begin{equation} \label{MEASURE-1}
 {\mathcal{N}} = \max_{\rho_{1,2}(0)} \int_{\sigma_{t} > 0}
 dt \; \sigma_{t}(t,\rho_{1,2}(0)),
\end{equation}
where $\sigma_{t}=\partial_{t}tr\Vert\rho_{1}(t)-\rho_{2}(t)\Vert/2$ and $\rho_{1,2}(0)$ is a pair of initial states.
For the damped Jaynes-Cummings model, we have $\sigma_{t}=\partial_{t}\rho_{11}$. We find that the non-Markovian effects increase the local velocity and further speed up the information back-flow from the environment to the system.  When $\gamma_{0}\gg\lambda_{0}$, the maximal local velocity means the maximal rate of the information back-flow, i.e., $V_{\max}=|\sigma_{t}|_{\max}$.
\bibliography{reference}

\begin{thebibliography}{67}%
\makeatletter
\providecommand \@ifxundefined [1]{%
 \@ifx{#1\undefined}
}%
\providecommand \@ifnum [1]{%
 \ifnum #1\expandafter \@firstoftwo
 \else \expandafter \@secondoftwo
 \fi
}%
\providecommand \@ifx [1]{%
 \ifx #1\expandafter \@firstoftwo
 \else \expandafter \@secondoftwo
 \fi
}%
\providecommand \natexlab [1]{#1}%
\providecommand \enquote  [1]{``#1''}%
\providecommand \bibnamefont  [1]{#1}%
\providecommand \bibfnamefont [1]{#1}%
\providecommand \citenamefont [1]{#1}%
\providecommand \href@noop [0]{\@secondoftwo}%
\providecommand \href [0]{\begingroup \@sanitize@url \@href}%
\providecommand \@href[1]{\@@startlink{#1}\@@href}%
\providecommand \@@href[1]{\endgroup#1\@@endlink}%
\providecommand \@sanitize@url [0]{\catcode `\\12\catcode `\$12\catcode
  `\&12\catcode `\#12\catcode `\^12\catcode `\_12\catcode `\%12\relax}%
\providecommand \@@startlink[1]{}%
\providecommand \@@endlink[0]{}%
\providecommand \url  [0]{\begingroup\@sanitize@url \@url }%
\providecommand \@url [1]{\endgroup\@href {#1}{\urlprefix }}%
\providecommand \urlprefix  [0]{URL }%
\providecommand \Eprint [0]{\href }%
\providecommand \doibase [0]{http://dx.doi.org/}%
\providecommand \selectlanguage [0]{\@gobble}%
\providecommand \bibinfo  [0]{\@secondoftwo}%
\providecommand \bibfield  [0]{\@secondoftwo}%
\providecommand \translation [1]{[#1]}%
\providecommand \BibitemOpen [0]{}%
\providecommand \bibitemStop [0]{}%
\providecommand \bibitemNoStop [0]{.\EOS\space}%
\providecommand \EOS [0]{\spacefactor3000\relax}%
\providecommand \BibitemShut  [1]{\csname bibitem#1\endcsname}%
\let\auto@bib@innerbib\@empty
\bibitem [{\citenamefont {Lloyd}(2000)}]{lloyd2000ultimate}%
  \BibitemOpen
  \bibfield  {author} {\bibinfo {author} {\bibfnamefont {S.}~\bibnamefont
  {Lloyd}},\ }\href {\doibase https://doi.org/10.1038/35023282} {\bibfield
  {journal} {\bibinfo  {journal} {Nature}\ }\textbf {\bibinfo {volume} {406}},\
  \bibinfo {pages} {1047} (\bibinfo {year} {2000})}\BibitemShut {NoStop}%
\bibitem [{\citenamefont {Campaioli}\ \emph {et~al.}(2017)\citenamefont
  {Campaioli}, \citenamefont {Pollock}, \citenamefont {Binder}, \citenamefont
  {C\'eleri}, \citenamefont {Goold}, \citenamefont {Vinjanampathy},\ and\
  \citenamefont {Modi}}]{PhysRevLett.118.150601}%
  \BibitemOpen
  \bibfield  {author} {\bibinfo {author} {\bibfnamefont {F.}~\bibnamefont
  {Campaioli}}, \bibinfo {author} {\bibfnamefont {F.~A.}\ \bibnamefont
  {Pollock}}, \bibinfo {author} {\bibfnamefont {F.~C.}\ \bibnamefont {Binder}},
  \bibinfo {author} {\bibfnamefont {L.}~\bibnamefont {C\'eleri}}, \bibinfo
  {author} {\bibfnamefont {J.}~\bibnamefont {Goold}}, \bibinfo {author}
  {\bibfnamefont {S.}~\bibnamefont {Vinjanampathy}}, \ and\ \bibinfo {author}
  {\bibfnamefont {K.}~\bibnamefont {Modi}},\ }\href {\doibase
  10.1103/PhysRevLett.118.150601} {\bibfield  {journal} {\bibinfo  {journal}
  {Phys. Rev. Lett.}\ }\textbf {\bibinfo {volume} {118}},\ \bibinfo {pages}
  {150601} (\bibinfo {year} {2017})}\BibitemShut {NoStop}%
\bibitem [{\citenamefont {Caneva}\ \emph {et~al.}(2009)\citenamefont {Caneva},
  \citenamefont {Murphy}, \citenamefont {Calarco}, \citenamefont {Fazio},
  \citenamefont {Montangero}, \citenamefont {Giovannetti},\ and\ \citenamefont
  {Santoro}}]{PhysRevLett.103.240501}%
  \BibitemOpen
  \bibfield  {author} {\bibinfo {author} {\bibfnamefont {T.}~\bibnamefont
  {Caneva}}, \bibinfo {author} {\bibfnamefont {M.}~\bibnamefont {Murphy}},
  \bibinfo {author} {\bibfnamefont {T.}~\bibnamefont {Calarco}}, \bibinfo
  {author} {\bibfnamefont {R.}~\bibnamefont {Fazio}}, \bibinfo {author}
  {\bibfnamefont {S.}~\bibnamefont {Montangero}}, \bibinfo {author}
  {\bibfnamefont {V.}~\bibnamefont {Giovannetti}}, \ and\ \bibinfo {author}
  {\bibfnamefont {G.~E.}\ \bibnamefont {Santoro}},\ }\href {\doibase
  10.1103/PhysRevLett.103.240501} {\bibfield  {journal} {\bibinfo  {journal}
  {Phys. Rev. Lett.}\ }\textbf {\bibinfo {volume} {103}},\ \bibinfo {pages}
  {240501} (\bibinfo {year} {2009})}\BibitemShut {NoStop}%
\bibitem [{\citenamefont {Campbell}\ and\ \citenamefont
  {Deffner}(2017)}]{PhysRevLett.118.100601}%
  \BibitemOpen
  \bibfield  {author} {\bibinfo {author} {\bibfnamefont {S.}~\bibnamefont
  {Campbell}}\ and\ \bibinfo {author} {\bibfnamefont {S.}~\bibnamefont
  {Deffner}},\ }\href {\doibase 10.1103/PhysRevLett.118.100601} {\bibfield
  {journal} {\bibinfo  {journal} {Phys. Rev. Lett.}\ }\textbf {\bibinfo
  {volume} {118}},\ \bibinfo {pages} {100601} (\bibinfo {year}
  {2017})}\BibitemShut {NoStop}%
\bibitem [{\citenamefont {Beau}\ and\ \citenamefont {del
  Campo}(2017)}]{PhysRevLett.119.010403}%
  \BibitemOpen
  \bibfield  {author} {\bibinfo {author} {\bibfnamefont {M.}~\bibnamefont
  {Beau}}\ and\ \bibinfo {author} {\bibfnamefont {A.}~\bibnamefont {del
  Campo}},\ }\href {\doibase 10.1103/PhysRevLett.119.010403} {\bibfield
  {journal} {\bibinfo  {journal} {Phys. Rev. Lett.}\ }\textbf {\bibinfo
  {volume} {119}},\ \bibinfo {pages} {010403} (\bibinfo {year}
  {2017})}\BibitemShut {NoStop}%
\bibitem [{\citenamefont {Zwierz}\ \emph {et~al.}(2010)\citenamefont {Zwierz},
  \citenamefont {P\'erez-Delgado},\ and\ \citenamefont
  {Kok}}]{PhysRevLett.105.180402}%
  \BibitemOpen
  \bibfield  {author} {\bibinfo {author} {\bibfnamefont {M.}~\bibnamefont
  {Zwierz}}, \bibinfo {author} {\bibfnamefont {C.~A.}\ \bibnamefont
  {P\'erez-Delgado}}, \ and\ \bibinfo {author} {\bibfnamefont {P.}~\bibnamefont
  {Kok}},\ }\href {\doibase 10.1103/PhysRevLett.105.180402} {\bibfield
  {journal} {\bibinfo  {journal} {Phys. Rev. Lett.}\ }\textbf {\bibinfo
  {volume} {105}},\ \bibinfo {pages} {180402} (\bibinfo {year}
  {2010})}\BibitemShut {NoStop}%
\bibitem [{\citenamefont {Deffner}\ and\ \citenamefont
  {Campbell}(2017)}]{Deffner_2017}%
  \BibitemOpen
  \bibfield  {author} {\bibinfo {author} {\bibfnamefont {S.}~\bibnamefont
  {Deffner}}\ and\ \bibinfo {author} {\bibfnamefont {S.}~\bibnamefont
  {Campbell}},\ }\href {\doibase 10.1088/1751-8121/aa86c6} {\bibfield
  {journal} {\bibinfo  {journal} {J. Phys. A: Math. Theor.}\ }\textbf {\bibinfo
  {volume} {50}},\ \bibinfo {pages} {453001} (\bibinfo {year}
  {2017})}\BibitemShut {NoStop}%
\bibitem [{\citenamefont {Frey}(2016)}]{WOS:000383587100001}%
  \BibitemOpen
  \bibfield  {author} {\bibinfo {author} {\bibfnamefont {M.~R.}\ \bibnamefont
  {Frey}},\ }\href {\doibase 10.1007/s11128-016-1405-x} {\bibfield  {journal}
  {\bibinfo  {journal} {Quantum Inf. Process.}\ }\textbf {\bibinfo {volume}
  {15}},\ \bibinfo {pages} {3919} (\bibinfo {year} {2016})}\BibitemShut
  {NoStop}%
\bibitem [{\citenamefont {Mandelstam}\ and\ \citenamefont
  {Tamm}(1945)}]{mandelstam1945uncertainty}%
  \BibitemOpen
  \bibfield  {author} {\bibinfo {author} {\bibfnamefont {L.}~\bibnamefont
  {Mandelstam}}\ and\ \bibinfo {author} {\bibfnamefont {I.}~\bibnamefont
  {Tamm}},\ }\href@noop {} {\bibfield  {journal} {\bibinfo  {journal} {J. Phys.
  (USSR)}\ }\textbf {\bibinfo {volume} {9}},\ \bibinfo {pages} {249} (\bibinfo
  {year} {1945})}\BibitemShut {NoStop}%
\bibitem [{\citenamefont {Margolus}\ and\ \citenamefont
  {Levitin}(1998)}]{MARGOLUS1998188}%
  \BibitemOpen
  \bibfield  {author} {\bibinfo {author} {\bibfnamefont {N.}~\bibnamefont
  {Margolus}}\ and\ \bibinfo {author} {\bibfnamefont {L.~B.}\ \bibnamefont
  {Levitin}},\ }\href {\doibase https://doi.org/10.1016/S0167-2789(98)00054-2}
  {\bibfield  {journal} {\bibinfo  {journal} {Physica D}\ }\textbf {\bibinfo
  {volume} {120}},\ \bibinfo {pages} {188} (\bibinfo {year}
  {1998})}\BibitemShut {NoStop}%
\bibitem [{\citenamefont {Anandan}\ and\ \citenamefont
  {Aharonov}(1990)}]{PhysRevLett.65.1697}%
  \BibitemOpen
  \bibfield  {author} {\bibinfo {author} {\bibfnamefont {J.}~\bibnamefont
  {Anandan}}\ and\ \bibinfo {author} {\bibfnamefont {Y.}~\bibnamefont
  {Aharonov}},\ }\href {\doibase 10.1103/PhysRevLett.65.1697} {\bibfield
  {journal} {\bibinfo  {journal} {Phys. Rev. Lett.}\ }\textbf {\bibinfo
  {volume} {65}},\ \bibinfo {pages} {1697} (\bibinfo {year}
  {1990})}\BibitemShut {NoStop}%
\bibitem [{\citenamefont {Cai}\ and\ \citenamefont
  {Zheng}(2017)}]{PhysRevA.95.052104}%
  \BibitemOpen
  \bibfield  {author} {\bibinfo {author} {\bibfnamefont {X.}~\bibnamefont
  {Cai}}\ and\ \bibinfo {author} {\bibfnamefont {Y.}~\bibnamefont {Zheng}},\
  }\href {\doibase 10.1103/PhysRevA.95.052104} {\bibfield  {journal} {\bibinfo
  {journal} {Phys. Rev. A}\ }\textbf {\bibinfo {volume} {95}},\ \bibinfo
  {pages} {052104} (\bibinfo {year} {2017})}\BibitemShut {NoStop}%
\bibitem [{\citenamefont {Taddei}\ \emph {et~al.}(2013)\citenamefont {Taddei},
  \citenamefont {Escher}, \citenamefont {Davidovich},\ and\ \citenamefont
  {de~Matos~Filho}}]{PhysRevLett.110.050402}%
  \BibitemOpen
  \bibfield  {author} {\bibinfo {author} {\bibfnamefont {M.~M.}\ \bibnamefont
  {Taddei}}, \bibinfo {author} {\bibfnamefont {B.~M.}\ \bibnamefont {Escher}},
  \bibinfo {author} {\bibfnamefont {L.}~\bibnamefont {Davidovich}}, \ and\
  \bibinfo {author} {\bibfnamefont {R.~L.}\ \bibnamefont {de~Matos~Filho}},\
  }\href {\doibase 10.1103/PhysRevLett.110.050402} {\bibfield  {journal}
  {\bibinfo  {journal} {Phys. Rev. Lett.}\ }\textbf {\bibinfo {volume} {110}},\
  \bibinfo {pages} {050402} (\bibinfo {year} {2013})}\BibitemShut {NoStop}%
\bibitem [{\citenamefont {Deffner}\ and\ \citenamefont
  {Lutz}(2013)}]{PhysRevLett.111.010402}%
  \BibitemOpen
  \bibfield  {author} {\bibinfo {author} {\bibfnamefont {S.}~\bibnamefont
  {Deffner}}\ and\ \bibinfo {author} {\bibfnamefont {E.}~\bibnamefont {Lutz}},\
  }\href {\doibase 10.1103/PhysRevLett.111.010402} {\bibfield  {journal}
  {\bibinfo  {journal} {Phys. Rev. Lett.}\ }\textbf {\bibinfo {volume} {111}},\
  \bibinfo {pages} {010402} (\bibinfo {year} {2013})}\BibitemShut {NoStop}%
\bibitem [{\citenamefont {del Campo}(2021)}]{PhysRevLett.126.180603}%
  \BibitemOpen
  \bibfield  {author} {\bibinfo {author} {\bibfnamefont {A.}~\bibnamefont {del
  Campo}},\ }\href {\doibase 10.1103/PhysRevLett.126.180603} {\bibfield
  {journal} {\bibinfo  {journal} {Phys. Rev. Lett.}\ }\textbf {\bibinfo
  {volume} {126}},\ \bibinfo {pages} {180603} (\bibinfo {year}
  {2021})}\BibitemShut {NoStop}%
\bibitem [{\citenamefont {del Campo}\ \emph {et~al.}(2013)\citenamefont {del
  Campo}, \citenamefont {Egusquiza}, \citenamefont {Plenio},\ and\
  \citenamefont {Huelga}}]{PhysRevLett.110.050403}%
  \BibitemOpen
  \bibfield  {author} {\bibinfo {author} {\bibfnamefont {A.}~\bibnamefont {del
  Campo}}, \bibinfo {author} {\bibfnamefont {I.~L.}\ \bibnamefont {Egusquiza}},
  \bibinfo {author} {\bibfnamefont {M.~B.}\ \bibnamefont {Plenio}}, \ and\
  \bibinfo {author} {\bibfnamefont {S.~F.}\ \bibnamefont {Huelga}},\ }\href
  {\doibase 10.1103/PhysRevLett.110.050403} {\bibfield  {journal} {\bibinfo
  {journal} {Phys. Rev. Lett.}\ }\textbf {\bibinfo {volume} {110}},\ \bibinfo
  {pages} {050403} (\bibinfo {year} {2013})}\BibitemShut {NoStop}%
\bibitem [{\citenamefont {Meng}\ \emph {et~al.}(2015)\citenamefont {Meng},
  \citenamefont {Wu},\ and\ \citenamefont {Guo}}]{Meng2015-sc}%
  \BibitemOpen
  \bibfield  {author} {\bibinfo {author} {\bibfnamefont {X.}~\bibnamefont
  {Meng}}, \bibinfo {author} {\bibfnamefont {C.}~\bibnamefont {Wu}}, \ and\
  \bibinfo {author} {\bibfnamefont {H.}~\bibnamefont {Guo}},\ }\href
  {https://doi.org/10.1038/srep16357} {\bibfield  {journal} {\bibinfo
  {journal} {Sci. Rep.}\ }\textbf {\bibinfo {volume} {5}},\ \bibinfo {pages}
  {16357} (\bibinfo {year} {2015})}\BibitemShut {NoStop}%
\bibitem [{\citenamefont {Mondal}\ \emph {et~al.}(2016)\citenamefont {Mondal},
  \citenamefont {Datta},\ and\ \citenamefont {Sazim}}]{MONDAL2016689}%
  \BibitemOpen
  \bibfield  {author} {\bibinfo {author} {\bibfnamefont {D.}~\bibnamefont
  {Mondal}}, \bibinfo {author} {\bibfnamefont {C.}~\bibnamefont {Datta}}, \
  and\ \bibinfo {author} {\bibfnamefont {S.}~\bibnamefont {Sazim}},\ }\href
  {\doibase https://doi.org/10.1016/j.physleta.2015.12.015} {\bibfield
  {journal} {\bibinfo  {journal} {Phys. Lett. A}\ }\textbf {\bibinfo {volume}
  {380}},\ \bibinfo {pages} {689} (\bibinfo {year} {2016})}\BibitemShut
  {NoStop}%
\bibitem [{\citenamefont {Campaioli}\ \emph {et~al.}(2018)\citenamefont
  {Campaioli}, \citenamefont {Pollock}, \citenamefont {Binder},\ and\
  \citenamefont {Modi}}]{PhysRevLett.120.060409}%
  \BibitemOpen
  \bibfield  {author} {\bibinfo {author} {\bibfnamefont {F.}~\bibnamefont
  {Campaioli}}, \bibinfo {author} {\bibfnamefont {F.~A.}\ \bibnamefont
  {Pollock}}, \bibinfo {author} {\bibfnamefont {F.~C.}\ \bibnamefont {Binder}},
  \ and\ \bibinfo {author} {\bibfnamefont {K.}~\bibnamefont {Modi}},\ }\href
  {\doibase 10.1103/PhysRevLett.120.060409} {\bibfield  {journal} {\bibinfo
  {journal} {Phys. Rev. Lett.}\ }\textbf {\bibinfo {volume} {120}},\ \bibinfo
  {pages} {060409} (\bibinfo {year} {2018})}\BibitemShut {NoStop}%
\bibitem [{\citenamefont {Marvian}\ and\ \citenamefont
  {Lidar}(2015)}]{PhysRevLett.115.210402}%
  \BibitemOpen
  \bibfield  {author} {\bibinfo {author} {\bibfnamefont {I.}~\bibnamefont
  {Marvian}}\ and\ \bibinfo {author} {\bibfnamefont {D.~A.}\ \bibnamefont
  {Lidar}},\ }\href {\doibase 10.1103/PhysRevLett.115.210402} {\bibfield
  {journal} {\bibinfo  {journal} {Phys. Rev. Lett.}\ }\textbf {\bibinfo
  {volume} {115}},\ \bibinfo {pages} {210402} (\bibinfo {year}
  {2015})}\BibitemShut {NoStop}%
\bibitem [{\citenamefont {Ness}\ \emph {et~al.}(2022)\citenamefont {Ness},
  \citenamefont {Alberti},\ and\ \citenamefont
  {Sagi}}]{PhysRevLett.129.140403}%
  \BibitemOpen
  \bibfield  {author} {\bibinfo {author} {\bibfnamefont {G.}~\bibnamefont
  {Ness}}, \bibinfo {author} {\bibfnamefont {A.}~\bibnamefont {Alberti}}, \
  and\ \bibinfo {author} {\bibfnamefont {Y.}~\bibnamefont {Sagi}},\ }\href
  {\doibase 10.1103/PhysRevLett.129.140403} {\bibfield  {journal} {\bibinfo
  {journal} {Phys. Rev. Lett.}\ }\textbf {\bibinfo {volume} {129}},\ \bibinfo
  {pages} {140403} (\bibinfo {year} {2022})}\BibitemShut {NoStop}%
\bibitem [{\citenamefont {Sun}\ and\ \citenamefont
  {Zheng}(2019)}]{PhysRevLett.123.180403}%
  \BibitemOpen
  \bibfield  {author} {\bibinfo {author} {\bibfnamefont {S.}~\bibnamefont
  {Sun}}\ and\ \bibinfo {author} {\bibfnamefont {Y.}~\bibnamefont {Zheng}},\
  }\href {\doibase 10.1103/PhysRevLett.123.180403} {\bibfield  {journal}
  {\bibinfo  {journal} {Phys. Rev. Lett.}\ }\textbf {\bibinfo {volume} {123}},\
  \bibinfo {pages} {180403} (\bibinfo {year} {2019})}\BibitemShut {NoStop}%
\bibitem [{\citenamefont {Xu}\ \emph {et~al.}(2020)\citenamefont {Xu},
  \citenamefont {Li}, \citenamefont {Busch}, \citenamefont {Chen},\ and\
  \citenamefont {Fogarty}}]{PhysRevResearch.2.023125}%
  \BibitemOpen
  \bibfield  {author} {\bibinfo {author} {\bibfnamefont {T.-N.}\ \bibnamefont
  {Xu}}, \bibinfo {author} {\bibfnamefont {J.}~\bibnamefont {Li}}, \bibinfo
  {author} {\bibfnamefont {T.}~\bibnamefont {Busch}}, \bibinfo {author}
  {\bibfnamefont {X.}~\bibnamefont {Chen}}, \ and\ \bibinfo {author}
  {\bibfnamefont {T.}~\bibnamefont {Fogarty}},\ }\href {\doibase
  10.1103/PhysRevResearch.2.023125} {\bibfield  {journal} {\bibinfo  {journal}
  {Phys. Rev. Research}\ }\textbf {\bibinfo {volume} {2}},\ \bibinfo {pages}
  {023125} (\bibinfo {year} {2020})}\BibitemShut {NoStop}%
\bibitem [{\citenamefont {Suzuki}\ and\ \citenamefont
  {Takahashi}(2020)}]{PhysRevResearch.2.032016}%
  \BibitemOpen
  \bibfield  {author} {\bibinfo {author} {\bibfnamefont {K.}~\bibnamefont
  {Suzuki}}\ and\ \bibinfo {author} {\bibfnamefont {K.}~\bibnamefont
  {Takahashi}},\ }\href {\doibase 10.1103/PhysRevResearch.2.032016} {\bibfield
  {journal} {\bibinfo  {journal} {Phys. Rev. Research}\ }\textbf {\bibinfo
  {volume} {2}},\ \bibinfo {pages} {032016(R)} (\bibinfo {year}
  {2020})}\BibitemShut {NoStop}%
\bibitem [{\citenamefont {Bukov}\ \emph {et~al.}(2019)\citenamefont {Bukov},
  \citenamefont {Sels},\ and\ \citenamefont {Polkovnikov}}]{PhysRevX.9.011034}%
  \BibitemOpen
  \bibfield  {author} {\bibinfo {author} {\bibfnamefont {M.}~\bibnamefont
  {Bukov}}, \bibinfo {author} {\bibfnamefont {D.}~\bibnamefont {Sels}}, \ and\
  \bibinfo {author} {\bibfnamefont {A.}~\bibnamefont {Polkovnikov}},\ }\href
  {\doibase 10.1103/PhysRevX.9.011034} {\bibfield  {journal} {\bibinfo
  {journal} {Phys. Rev. X}\ }\textbf {\bibinfo {volume} {9}},\ \bibinfo {pages}
  {011034} (\bibinfo {year} {2019})}\BibitemShut {NoStop}%
\bibitem [{\citenamefont {Sun}\ \emph {et~al.}(2021)\citenamefont {Sun},
  \citenamefont {Peng}, \citenamefont {Hu},\ and\ \citenamefont
  {Zheng}}]{PhysRevLett.127.100404}%
  \BibitemOpen
  \bibfield  {author} {\bibinfo {author} {\bibfnamefont {S.}~\bibnamefont
  {Sun}}, \bibinfo {author} {\bibfnamefont {Y.}~\bibnamefont {Peng}}, \bibinfo
  {author} {\bibfnamefont {X.}~\bibnamefont {Hu}}, \ and\ \bibinfo {author}
  {\bibfnamefont {Y.}~\bibnamefont {Zheng}},\ }\href {\doibase
  10.1103/PhysRevLett.127.100404} {\bibfield  {journal} {\bibinfo  {journal}
  {Phys. Rev. Lett.}\ }\textbf {\bibinfo {volume} {127}},\ \bibinfo {pages}
  {100404} (\bibinfo {year} {2021})}\BibitemShut {NoStop}%
\bibitem [{\citenamefont {{Wu}}\ and\ \citenamefont
  {{An}}()}]{2022arXiv220702438W}%
  \BibitemOpen
  \bibfield  {author} {\bibinfo {author} {\bibfnamefont {W.}~\bibnamefont
  {{Wu}}}\ and\ \bibinfo {author} {\bibfnamefont {J.-H.}\ \bibnamefont
  {{An}}},\ }\href@noop {} {\ }\Eprint {http://arxiv.org/abs/2207.02438}
  {arXiv:2207.02438} \BibitemShut {NoStop}%
\bibitem [{\citenamefont {Pires}\ \emph {et~al.}(2016)\citenamefont {Pires},
  \citenamefont {Cianciaruso}, \citenamefont {C\'eleri}, \citenamefont
  {Adesso},\ and\ \citenamefont {Soares-Pinto}}]{PhysRevX.6.021031}%
  \BibitemOpen
  \bibfield  {author} {\bibinfo {author} {\bibfnamefont {D.~P.}\ \bibnamefont
  {Pires}}, \bibinfo {author} {\bibfnamefont {M.}~\bibnamefont {Cianciaruso}},
  \bibinfo {author} {\bibfnamefont {L.~C.}\ \bibnamefont {C\'eleri}}, \bibinfo
  {author} {\bibfnamefont {G.}~\bibnamefont {Adesso}}, \ and\ \bibinfo {author}
  {\bibfnamefont {D.~O.}\ \bibnamefont {Soares-Pinto}},\ }\href {\doibase
  10.1103/PhysRevX.6.021031} {\bibfield  {journal} {\bibinfo  {journal} {Phys.
  Rev. X}\ }\textbf {\bibinfo {volume} {6}},\ \bibinfo {pages} {021031}
  (\bibinfo {year} {2016})}\BibitemShut {NoStop}%
\bibitem [{\citenamefont {Garc\'{\i}a-Pintos}\ \emph
  {et~al.}(2022)\citenamefont {Garc\'{\i}a-Pintos}, \citenamefont {Nicholson},
  \citenamefont {Green}, \citenamefont {del Campo},\ and\ \citenamefont
  {Gorshkov}}]{PhysRevX.12.011038}%
  \BibitemOpen
  \bibfield  {author} {\bibinfo {author} {\bibfnamefont {L.~P.}\ \bibnamefont
  {Garc\'{\i}a-Pintos}}, \bibinfo {author} {\bibfnamefont {S.~B.}\ \bibnamefont
  {Nicholson}}, \bibinfo {author} {\bibfnamefont {J.~R.}\ \bibnamefont
  {Green}}, \bibinfo {author} {\bibfnamefont {A.}~\bibnamefont {del Campo}}, \
  and\ \bibinfo {author} {\bibfnamefont {A.~V.}\ \bibnamefont {Gorshkov}},\
  }\href {\doibase 10.1103/PhysRevX.12.011038} {\bibfield  {journal} {\bibinfo
  {journal} {Phys. Rev. X}\ }\textbf {\bibinfo {volume} {12}},\ \bibinfo
  {pages} {011038} (\bibinfo {year} {2022})}\BibitemShut {NoStop}%
\bibitem [{\citenamefont {Poggi}\ \emph {et~al.}(2021)\citenamefont {Poggi},
  \citenamefont {Campbell},\ and\ \citenamefont
  {Deffner}}]{PRXQuantum.2.040349}%
  \BibitemOpen
  \bibfield  {author} {\bibinfo {author} {\bibfnamefont {P.~M.}\ \bibnamefont
  {Poggi}}, \bibinfo {author} {\bibfnamefont {S.}~\bibnamefont {Campbell}}, \
  and\ \bibinfo {author} {\bibfnamefont {S.}~\bibnamefont {Deffner}},\ }\href
  {\doibase 10.1103/PRXQuantum.2.040349} {\bibfield  {journal} {\bibinfo
  {journal} {PRX Quantum}\ }\textbf {\bibinfo {volume} {2}},\ \bibinfo {pages}
  {040349} (\bibinfo {year} {2021})}\BibitemShut {NoStop}%
\bibitem [{\citenamefont {Okuyama}\ and\ \citenamefont
  {Ohzeki}(2018)}]{PhysRevLett.120.070402}%
  \BibitemOpen
  \bibfield  {author} {\bibinfo {author} {\bibfnamefont {M.}~\bibnamefont
  {Okuyama}}\ and\ \bibinfo {author} {\bibfnamefont {M.}~\bibnamefont
  {Ohzeki}},\ }\href {\doibase 10.1103/PhysRevLett.120.070402} {\bibfield
  {journal} {\bibinfo  {journal} {Phys. Rev. Lett.}\ }\textbf {\bibinfo
  {volume} {120}},\ \bibinfo {pages} {070402} (\bibinfo {year}
  {2018})}\BibitemShut {NoStop}%
\bibitem [{\citenamefont {Shanahan}\ \emph {et~al.}(2018)\citenamefont
  {Shanahan}, \citenamefont {Chenu}, \citenamefont {Margolus},\ and\
  \citenamefont {del Campo}}]{PhysRevLett.120.070401}%
  \BibitemOpen
  \bibfield  {author} {\bibinfo {author} {\bibfnamefont {B.}~\bibnamefont
  {Shanahan}}, \bibinfo {author} {\bibfnamefont {A.}~\bibnamefont {Chenu}},
  \bibinfo {author} {\bibfnamefont {N.}~\bibnamefont {Margolus}}, \ and\
  \bibinfo {author} {\bibfnamefont {A.}~\bibnamefont {del Campo}},\ }\href
  {\doibase 10.1103/PhysRevLett.120.070401} {\bibfield  {journal} {\bibinfo
  {journal} {Phys. Rev. Lett.}\ }\textbf {\bibinfo {volume} {120}},\ \bibinfo
  {pages} {070401} (\bibinfo {year} {2018})}\BibitemShut {NoStop}%
\bibitem [{\citenamefont {Van~Vu}\ and\ \citenamefont
  {Saito}(2023)}]{PhysRevLett.130.010402}%
  \BibitemOpen
  \bibfield  {author} {\bibinfo {author} {\bibfnamefont {T.}~\bibnamefont
  {Van~Vu}}\ and\ \bibinfo {author} {\bibfnamefont {K.}~\bibnamefont {Saito}},\
  }\href {\doibase 10.1103/PhysRevLett.130.010402} {\bibfield  {journal}
  {\bibinfo  {journal} {Phys. Rev. Lett.}\ }\textbf {\bibinfo {volume} {130}},\
  \bibinfo {pages} {010402} (\bibinfo {year} {2023})}\BibitemShut {NoStop}%
\bibitem [{\citenamefont {Carlini}\ \emph {et~al.}(2006)\citenamefont
  {Carlini}, \citenamefont {Hosoya}, \citenamefont {Koike},\ and\ \citenamefont
  {Okudaira}}]{PhysRevLett.96.060503}%
  \BibitemOpen
  \bibfield  {author} {\bibinfo {author} {\bibfnamefont {A.}~\bibnamefont
  {Carlini}}, \bibinfo {author} {\bibfnamefont {A.}~\bibnamefont {Hosoya}},
  \bibinfo {author} {\bibfnamefont {T.}~\bibnamefont {Koike}}, \ and\ \bibinfo
  {author} {\bibfnamefont {Y.}~\bibnamefont {Okudaira}},\ }\href {\doibase
  10.1103/PhysRevLett.96.060503} {\bibfield  {journal} {\bibinfo  {journal}
  {Phys. Rev. Lett.}\ }\textbf {\bibinfo {volume} {96}},\ \bibinfo {pages}
  {060503} (\bibinfo {year} {2006})}\BibitemShut {NoStop}%
\bibitem [{\citenamefont {Wang}\ \emph {et~al.}(2015)\citenamefont {Wang},
  \citenamefont {Allegra}, \citenamefont {Jacobs}, \citenamefont {Lloyd},
  \citenamefont {Lupo},\ and\ \citenamefont
  {Mohseni}}]{PhysRevLett.114.170501}%
  \BibitemOpen
  \bibfield  {author} {\bibinfo {author} {\bibfnamefont {X.}~\bibnamefont
  {Wang}}, \bibinfo {author} {\bibfnamefont {M.}~\bibnamefont {Allegra}},
  \bibinfo {author} {\bibfnamefont {K.}~\bibnamefont {Jacobs}}, \bibinfo
  {author} {\bibfnamefont {S.}~\bibnamefont {Lloyd}}, \bibinfo {author}
  {\bibfnamefont {C.}~\bibnamefont {Lupo}}, \ and\ \bibinfo {author}
  {\bibfnamefont {M.}~\bibnamefont {Mohseni}},\ }\href {\doibase
  10.1103/PhysRevLett.114.170501} {\bibfield  {journal} {\bibinfo  {journal}
  {Phys. Rev. Lett.}\ }\textbf {\bibinfo {volume} {114}},\ \bibinfo {pages}
  {170501} (\bibinfo {year} {2015})}\BibitemShut {NoStop}%
\bibitem [{\citenamefont {Koike}(2022)}]{WOS:000880809900002}%
  \BibitemOpen
  \bibfield  {author} {\bibinfo {author} {\bibfnamefont {T.}~\bibnamefont
  {Koike}},\ }\href {\doibase 10.1098/rsta.2021.0273} {\bibfield  {journal}
  {\bibinfo  {journal} {Philos. Trans. R. Soc., A}\ }\textbf {\bibinfo {volume}
  {380}},\ \bibinfo {pages} {20210273} (\bibinfo {year} {2022})}\BibitemShut
  {NoStop}%
\bibitem [{\citenamefont {Lam}\ \emph {et~al.}(2021)\citenamefont {Lam},
  \citenamefont {Peter}, \citenamefont {Groh}, \citenamefont {Alt},
  \citenamefont {Robens}, \citenamefont {Meschede}, \citenamefont {Negretti},
  \citenamefont {Montangero}, \citenamefont {Calarco},\ and\ \citenamefont
  {Alberti}}]{PhysRevX.11.011035}%
  \BibitemOpen
  \bibfield  {author} {\bibinfo {author} {\bibfnamefont {M.~R.}\ \bibnamefont
  {Lam}}, \bibinfo {author} {\bibfnamefont {N.}~\bibnamefont {Peter}}, \bibinfo
  {author} {\bibfnamefont {T.}~\bibnamefont {Groh}}, \bibinfo {author}
  {\bibfnamefont {W.}~\bibnamefont {Alt}}, \bibinfo {author} {\bibfnamefont
  {C.}~\bibnamefont {Robens}}, \bibinfo {author} {\bibfnamefont
  {D.}~\bibnamefont {Meschede}}, \bibinfo {author} {\bibfnamefont
  {A.}~\bibnamefont {Negretti}}, \bibinfo {author} {\bibfnamefont
  {S.}~\bibnamefont {Montangero}}, \bibinfo {author} {\bibfnamefont
  {T.}~\bibnamefont {Calarco}}, \ and\ \bibinfo {author} {\bibfnamefont
  {A.}~\bibnamefont {Alberti}},\ }\href {\doibase 10.1103/PhysRevX.11.011035}
  {\bibfield  {journal} {\bibinfo  {journal} {Phys. Rev. X}\ }\textbf {\bibinfo
  {volume} {11}},\ \bibinfo {pages} {011035} (\bibinfo {year}
  {2021})}\BibitemShut {NoStop}%
\bibitem [{\citenamefont {Kuzmak}\ and\ \citenamefont
  {Tkachuk}(2013)}]{Kuzmak_2013}%
  \BibitemOpen
  \bibfield  {author} {\bibinfo {author} {\bibfnamefont {A.~R.}\ \bibnamefont
  {Kuzmak}}\ and\ \bibinfo {author} {\bibfnamefont {V.~M.}\ \bibnamefont
  {Tkachuk}},\ }\href {\doibase 10.1088/1751-8113/46/15/155305} {\bibfield
  {journal} {\bibinfo  {journal} {J. Phys. A: Math. Theor.}\ }\textbf {\bibinfo
  {volume} {46}},\ \bibinfo {pages} {155305} (\bibinfo {year}
  {2013})}\BibitemShut {NoStop}%
\bibitem [{\citenamefont {Mostafazadeh}(2007)}]{PhysRevLett.99.130502}%
  \BibitemOpen
  \bibfield  {author} {\bibinfo {author} {\bibfnamefont {A.}~\bibnamefont
  {Mostafazadeh}},\ }\href {\doibase 10.1103/PhysRevLett.99.130502} {\bibfield
  {journal} {\bibinfo  {journal} {Phys. Rev. Lett.}\ }\textbf {\bibinfo
  {volume} {99}},\ \bibinfo {pages} {130502} (\bibinfo {year}
  {2007})}\BibitemShut {NoStop}%
\bibitem [{\citenamefont {G\"unther}\ and\ \citenamefont
  {Samsonov}(2008)}]{PhysRevLett.101.230404}%
  \BibitemOpen
  \bibfield  {author} {\bibinfo {author} {\bibfnamefont {U.}~\bibnamefont
  {G\"unther}}\ and\ \bibinfo {author} {\bibfnamefont {B.~F.}\ \bibnamefont
  {Samsonov}},\ }\href {\doibase 10.1103/PhysRevLett.101.230404} {\bibfield
  {journal} {\bibinfo  {journal} {Phys. Rev. Lett.}\ }\textbf {\bibinfo
  {volume} {101}},\ \bibinfo {pages} {230404} (\bibinfo {year}
  {2008})}\BibitemShut {NoStop}%
\bibitem [{\citenamefont {Albash}\ and\ \citenamefont
  {Lidar}(2018)}]{RevModPhys.90.015002}%
  \BibitemOpen
  \bibfield  {author} {\bibinfo {author} {\bibfnamefont {T.}~\bibnamefont
  {Albash}}\ and\ \bibinfo {author} {\bibfnamefont {D.~A.}\ \bibnamefont
  {Lidar}},\ }\href {\doibase 10.1103/RevModPhys.90.015002} {\bibfield
  {journal} {\bibinfo  {journal} {Rev. Mod. Phys.}\ }\textbf {\bibinfo {volume}
  {90}},\ \bibinfo {pages} {015002} (\bibinfo {year} {2018})}\BibitemShut
  {NoStop}%
\bibitem [{\citenamefont {Rezakhani}\ \emph {et~al.}(2009)\citenamefont
  {Rezakhani}, \citenamefont {Kuo}, \citenamefont {Hamma}, \citenamefont
  {Lidar},\ and\ \citenamefont {Zanardi}}]{PhysRevLett.103.080502}%
  \BibitemOpen
  \bibfield  {author} {\bibinfo {author} {\bibfnamefont {A.~T.}\ \bibnamefont
  {Rezakhani}}, \bibinfo {author} {\bibfnamefont {W.-J.}\ \bibnamefont {Kuo}},
  \bibinfo {author} {\bibfnamefont {A.}~\bibnamefont {Hamma}}, \bibinfo
  {author} {\bibfnamefont {D.~A.}\ \bibnamefont {Lidar}}, \ and\ \bibinfo
  {author} {\bibfnamefont {P.}~\bibnamefont {Zanardi}},\ }\href {\doibase
  10.1103/PhysRevLett.103.080502} {\bibfield  {journal} {\bibinfo  {journal}
  {Phys. Rev. Lett.}\ }\textbf {\bibinfo {volume} {103}},\ \bibinfo {pages}
  {080502} (\bibinfo {year} {2009})}\BibitemShut {NoStop}%
\bibitem [{\citenamefont {Campaioli}\ \emph {et~al.}(2019)\citenamefont
  {Campaioli}, \citenamefont {Sloan}, \citenamefont {Modi},\ and\ \citenamefont
  {Pollock}}]{PhysRevA.100.062328}%
  \BibitemOpen
  \bibfield  {author} {\bibinfo {author} {\bibfnamefont {F.}~\bibnamefont
  {Campaioli}}, \bibinfo {author} {\bibfnamefont {W.}~\bibnamefont {Sloan}},
  \bibinfo {author} {\bibfnamefont {K.}~\bibnamefont {Modi}}, \ and\ \bibinfo
  {author} {\bibfnamefont {F.~A.}\ \bibnamefont {Pollock}},\ }\href {\doibase
  10.1103/PhysRevA.100.062328} {\bibfield  {journal} {\bibinfo  {journal}
  {Phys. Rev. A}\ }\textbf {\bibinfo {volume} {100}},\ \bibinfo {pages}
  {062328} (\bibinfo {year} {2019})}\BibitemShut {NoStop}%
\bibitem [{\citenamefont {Santos}\ \emph {et~al.}(2021)\citenamefont {Santos},
  \citenamefont {Villas-Boas},\ and\ \citenamefont
  {Bachelard}}]{PhysRevA.103.012206}%
  \BibitemOpen
  \bibfield  {author} {\bibinfo {author} {\bibfnamefont {A.~C.}\ \bibnamefont
  {Santos}}, \bibinfo {author} {\bibfnamefont {C.~J.}\ \bibnamefont
  {Villas-Boas}}, \ and\ \bibinfo {author} {\bibfnamefont {R.}~\bibnamefont
  {Bachelard}},\ }\href {\doibase 10.1103/PhysRevA.103.012206} {\bibfield
  {journal} {\bibinfo  {journal} {Phys. Rev. A}\ }\textbf {\bibinfo {volume}
  {103}},\ \bibinfo {pages} {012206} (\bibinfo {year} {2021})}\BibitemShut
  {NoStop}%
\bibitem [{\citenamefont {Hegerfeldt}(2013)}]{PhysRevLett.111.260501}%
  \BibitemOpen
  \bibfield  {author} {\bibinfo {author} {\bibfnamefont {G.~C.}\ \bibnamefont
  {Hegerfeldt}},\ }\href {\doibase 10.1103/PhysRevLett.111.260501} {\bibfield
  {journal} {\bibinfo  {journal} {Phys. Rev. Lett.}\ }\textbf {\bibinfo
  {volume} {111}},\ \bibinfo {pages} {260501} (\bibinfo {year}
  {2013})}\BibitemShut {NoStop}%
\bibitem [{\citenamefont {Bason}\ \emph {et~al.}(2012)\citenamefont {Bason},
  \citenamefont {Viteau}, \citenamefont {Malossi}, \citenamefont {Huillery},
  \citenamefont {Arimondo}, \citenamefont {Ciampini}, \citenamefont {Fazio},
  \citenamefont {Giovannetti}, \citenamefont {Mannella},\ and\ \citenamefont
  {Morsch}}]{WOS:000300403700020}%
  \BibitemOpen
  \bibfield  {author} {\bibinfo {author} {\bibfnamefont {M.~G.}\ \bibnamefont
  {Bason}}, \bibinfo {author} {\bibfnamefont {M.}~\bibnamefont {Viteau}},
  \bibinfo {author} {\bibfnamefont {N.}~\bibnamefont {Malossi}}, \bibinfo
  {author} {\bibfnamefont {P.}~\bibnamefont {Huillery}}, \bibinfo {author}
  {\bibfnamefont {E.}~\bibnamefont {Arimondo}}, \bibinfo {author}
  {\bibfnamefont {D.}~\bibnamefont {Ciampini}}, \bibinfo {author}
  {\bibfnamefont {R.}~\bibnamefont {Fazio}}, \bibinfo {author} {\bibfnamefont
  {V.}~\bibnamefont {Giovannetti}}, \bibinfo {author} {\bibfnamefont
  {R.}~\bibnamefont {Mannella}}, \ and\ \bibinfo {author} {\bibfnamefont
  {O.}~\bibnamefont {Morsch}},\ }\href {\doibase 10.1038/NPHYS2170} {\bibfield
  {journal} {\bibinfo  {journal} {Nat. Phys.}\ }\textbf {\bibinfo {volume}
  {8}},\ \bibinfo {pages} {147} (\bibinfo {year} {2012})}\BibitemShut {NoStop}%
\bibitem [{\citenamefont {Wootters}(1981)}]{PhysRevD.23.357}%
  \BibitemOpen
  \bibfield  {author} {\bibinfo {author} {\bibfnamefont {W.~K.}\ \bibnamefont
  {Wootters}},\ }\href {\doibase 10.1103/PhysRevD.23.357} {\bibfield  {journal}
  {\bibinfo  {journal} {Phys. Rev. D}\ }\textbf {\bibinfo {volume} {23}},\
  \bibinfo {pages} {357} (\bibinfo {year} {1981})}\BibitemShut {NoStop}%
\bibitem [{\citenamefont {Braunstein}\ and\ \citenamefont
  {Caves}(1994)}]{PhysRevLett.72.3439}%
  \BibitemOpen
  \bibfield  {author} {\bibinfo {author} {\bibfnamefont {S.~L.}\ \bibnamefont
  {Braunstein}}\ and\ \bibinfo {author} {\bibfnamefont {C.~M.}\ \bibnamefont
  {Caves}},\ }\href {\doibase 10.1103/PhysRevLett.72.3439} {\bibfield
  {journal} {\bibinfo  {journal} {Phys. Rev. Lett.}\ }\textbf {\bibinfo
  {volume} {72}},\ \bibinfo {pages} {3439} (\bibinfo {year}
  {1994})}\BibitemShut {NoStop}%
\bibitem [{\citenamefont {Bengtsson}\ and\ \citenamefont
  {Życzkowski}(2017)}]{bengtsson2017}%
  \BibitemOpen
  \bibfield  {author} {\bibinfo {author} {\bibfnamefont {I.}~\bibnamefont
  {Bengtsson}}\ and\ \bibinfo {author} {\bibfnamefont {K.}~\bibnamefont
  {Życzkowski}},\ }\href {\doibase 10.1017/9781139207010} {\emph {\bibinfo
  {title} {Geometry of Quantum States: An Introduction to Quantum
  Entanglement}}},\ \bibinfo {edition} {2nd}\ ed.\ (\bibinfo  {publisher}
  {Cambridge University Press},\ \bibinfo {year} {2017})\BibitemShut {NoStop}%
\bibitem [{\citenamefont {Wu}\ and\ \citenamefont
  {Niu}(2000)}]{PhysRevA.61.023402}%
  \BibitemOpen
  \bibfield  {author} {\bibinfo {author} {\bibfnamefont {B.}~\bibnamefont
  {Wu}}\ and\ \bibinfo {author} {\bibfnamefont {Q.}~\bibnamefont {Niu}},\
  }\href {\doibase 10.1103/PhysRevA.61.023402} {\bibfield  {journal} {\bibinfo
  {journal} {Phys. Rev. A}\ }\textbf {\bibinfo {volume} {61}},\ \bibinfo
  {pages} {023402} (\bibinfo {year} {2000})}\BibitemShut {NoStop}%
\bibitem [{\citenamefont {Liu}\ \emph {et~al.}(2002)\citenamefont {Liu},
  \citenamefont {Fu}, \citenamefont {Ou}, \citenamefont {Chen}, \citenamefont
  {Choi}, \citenamefont {Wu},\ and\ \citenamefont {Niu}}]{PhysRevA.66.023404}%
  \BibitemOpen
  \bibfield  {author} {\bibinfo {author} {\bibfnamefont {J.}~\bibnamefont
  {Liu}}, \bibinfo {author} {\bibfnamefont {L.}~\bibnamefont {Fu}}, \bibinfo
  {author} {\bibfnamefont {B.-Y.}\ \bibnamefont {Ou}}, \bibinfo {author}
  {\bibfnamefont {S.-G.}\ \bibnamefont {Chen}}, \bibinfo {author}
  {\bibfnamefont {D.-I.}\ \bibnamefont {Choi}}, \bibinfo {author}
  {\bibfnamefont {B.}~\bibnamefont {Wu}}, \ and\ \bibinfo {author}
  {\bibfnamefont {Q.}~\bibnamefont {Niu}},\ }\href {\doibase
  10.1103/PhysRevA.66.023404} {\bibfield  {journal} {\bibinfo  {journal} {Phys.
  Rev. A}\ }\textbf {\bibinfo {volume} {66}},\ \bibinfo {pages} {023404}
  (\bibinfo {year} {2002})}\BibitemShut {NoStop}%
\bibitem [{\citenamefont {Jona-Lasinio}\ \emph {et~al.}(2003)\citenamefont
  {Jona-Lasinio}, \citenamefont {Morsch}, \citenamefont {Cristiani},
  \citenamefont {Malossi}, \citenamefont {M\"uller}, \citenamefont {Courtade},
  \citenamefont {Anderlini},\ and\ \citenamefont
  {Arimondo}}]{PhysRevLett.91.230406}%
  \BibitemOpen
  \bibfield  {author} {\bibinfo {author} {\bibfnamefont {M.}~\bibnamefont
  {Jona-Lasinio}}, \bibinfo {author} {\bibfnamefont {O.}~\bibnamefont
  {Morsch}}, \bibinfo {author} {\bibfnamefont {M.}~\bibnamefont {Cristiani}},
  \bibinfo {author} {\bibfnamefont {N.}~\bibnamefont {Malossi}}, \bibinfo
  {author} {\bibfnamefont {J.~H.}\ \bibnamefont {M\"uller}}, \bibinfo {author}
  {\bibfnamefont {E.}~\bibnamefont {Courtade}}, \bibinfo {author}
  {\bibfnamefont {M.}~\bibnamefont {Anderlini}}, \ and\ \bibinfo {author}
  {\bibfnamefont {E.}~\bibnamefont {Arimondo}},\ }\href {\doibase
  10.1103/PhysRevLett.91.230406} {\bibfield  {journal} {\bibinfo  {journal}
  {Phys. Rev. Lett.}\ }\textbf {\bibinfo {volume} {91}},\ \bibinfo {pages}
  {230406} (\bibinfo {year} {2003})}\BibitemShut {NoStop}%
\bibitem [{\citenamefont {Dou}\ \emph {et~al.}(2014)\citenamefont {Dou},
  \citenamefont {Fu},\ and\ \citenamefont {Liu}}]{PhysRevA.89.012123}%
  \BibitemOpen
  \bibfield  {author} {\bibinfo {author} {\bibfnamefont {F.-Q.}\ \bibnamefont
  {Dou}}, \bibinfo {author} {\bibfnamefont {L.-B.}\ \bibnamefont {Fu}}, \ and\
  \bibinfo {author} {\bibfnamefont {J.}~\bibnamefont {Liu}},\ }\href {\doibase
  10.1103/PhysRevA.89.012123} {\bibfield  {journal} {\bibinfo  {journal} {Phys.
  Rev. A}\ }\textbf {\bibinfo {volume} {89}},\ \bibinfo {pages} {012123}
  (\bibinfo {year} {2014})}\BibitemShut {NoStop}%
\bibitem [{\citenamefont {Dou}\ \emph {et~al.}(2016)\citenamefont {Dou},
  \citenamefont {Cao}, \citenamefont {Liu},\ and\ \citenamefont
  {Fu}}]{PhysRevA.93.043419}%
  \BibitemOpen
  \bibfield  {author} {\bibinfo {author} {\bibfnamefont {F.-Q.}\ \bibnamefont
  {Dou}}, \bibinfo {author} {\bibfnamefont {H.}~\bibnamefont {Cao}}, \bibinfo
  {author} {\bibfnamefont {J.}~\bibnamefont {Liu}}, \ and\ \bibinfo {author}
  {\bibfnamefont {L.-B.}\ \bibnamefont {Fu}},\ }\href {\doibase
  10.1103/PhysRevA.93.043419} {\bibfield  {journal} {\bibinfo  {journal} {Phys.
  Rev. A}\ }\textbf {\bibinfo {volume} {93}},\ \bibinfo {pages} {043419}
  (\bibinfo {year} {2016})}\BibitemShut {NoStop}%
\bibitem [{\citenamefont {Dou}\ \emph {et~al.}(2018)\citenamefont {Dou},
  \citenamefont {Liu},\ and\ \citenamefont {Fu}}]{PhysRevA.98.022102}%
  \BibitemOpen
  \bibfield  {author} {\bibinfo {author} {\bibfnamefont {F.-Q.}\ \bibnamefont
  {Dou}}, \bibinfo {author} {\bibfnamefont {J.}~\bibnamefont {Liu}}, \ and\
  \bibinfo {author} {\bibfnamefont {L.-B.}\ \bibnamefont {Fu}},\ }\href
  {\doibase 10.1103/PhysRevA.98.022102} {\bibfield  {journal} {\bibinfo
  {journal} {Phys. Rev. A}\ }\textbf {\bibinfo {volume} {98}},\ \bibinfo
  {pages} {022102} (\bibinfo {year} {2018})}\BibitemShut {NoStop}%
\bibitem [{\citenamefont {Chen}\ \emph {et~al.}(2016)\citenamefont {Chen},
  \citenamefont {Ban},\ and\ \citenamefont {Hegerfeldt}}]{PhysRevA.94.023624}%
  \BibitemOpen
  \bibfield  {author} {\bibinfo {author} {\bibfnamefont {X.}~\bibnamefont
  {Chen}}, \bibinfo {author} {\bibfnamefont {Y.}~\bibnamefont {Ban}}, \ and\
  \bibinfo {author} {\bibfnamefont {G.~C.}\ \bibnamefont {Hegerfeldt}},\ }\href
  {\doibase 10.1103/PhysRevA.94.023624} {\bibfield  {journal} {\bibinfo
  {journal} {Phys. Rev. A}\ }\textbf {\bibinfo {volume} {94}},\ \bibinfo
  {pages} {023624} (\bibinfo {year} {2016})}\BibitemShut {NoStop}%
\bibitem [{\citenamefont {Berry}(2009)}]{Berry_2009}%
  \BibitemOpen
  \bibfield  {author} {\bibinfo {author} {\bibfnamefont {M.~V.}\ \bibnamefont
  {Berry}},\ }\href {\doibase 10.1088/1751-8113/42/36/365303} {\bibfield
  {journal} {\bibinfo  {journal} {J. Phys. A: Math. Theor.}\ }\textbf {\bibinfo
  {volume} {42}},\ \bibinfo {pages} {365303} (\bibinfo {year}
  {2009})}\BibitemShut {NoStop}%
\bibitem [{\citenamefont {Uhlmann}(1993)}]{UHLMANN1993253}%
  \BibitemOpen
  \bibfield  {author} {\bibinfo {author} {\bibfnamefont {A.}~\bibnamefont
  {Uhlmann}},\ }\href {\doibase https://doi.org/10.1016/0034-4877(93)90060-R}
  {\bibfield  {journal} {\bibinfo  {journal} {Rep. Math. Phys.}\ }\textbf
  {\bibinfo {volume} {33}},\ \bibinfo {pages} {253} (\bibinfo {year}
  {1993})}\BibitemShut {NoStop}%
\bibitem [{\citenamefont {Breuer}\ \emph {et~al.}(1999)\citenamefont {Breuer},
  \citenamefont {Kappler},\ and\ \citenamefont
  {Petruccione}}]{PhysRevA.59.1633}%
  \BibitemOpen
  \bibfield  {author} {\bibinfo {author} {\bibfnamefont {H.-P.}\ \bibnamefont
  {Breuer}}, \bibinfo {author} {\bibfnamefont {B.}~\bibnamefont {Kappler}}, \
  and\ \bibinfo {author} {\bibfnamefont {F.}~\bibnamefont {Petruccione}},\
  }\href {\doibase 10.1103/PhysRevA.59.1633} {\bibfield  {journal} {\bibinfo
  {journal} {Phys. Rev. A}\ }\textbf {\bibinfo {volume} {59}},\ \bibinfo
  {pages} {1633} (\bibinfo {year} {1999})}\BibitemShut {NoStop}%
\bibitem [{\citenamefont {Garraway}(1997)}]{PhysRevA.55.2290}%
  \BibitemOpen
  \bibfield  {author} {\bibinfo {author} {\bibfnamefont {B.~M.}\ \bibnamefont
  {Garraway}},\ }\href {\doibase 10.1103/PhysRevA.55.2290} {\bibfield
  {journal} {\bibinfo  {journal} {Phys. Rev. A}\ }\textbf {\bibinfo {volume}
  {55}},\ \bibinfo {pages} {2290} (\bibinfo {year} {1997})}\BibitemShut
  {NoStop}%
\bibitem [{\citenamefont {Breuer}\ \emph {et~al.}(2009)\citenamefont {Breuer},
  \citenamefont {Laine},\ and\ \citenamefont {Piilo}}]{PhysRevLett.103.210401}%
  \BibitemOpen
  \bibfield  {author} {\bibinfo {author} {\bibfnamefont {H.-P.}\ \bibnamefont
  {Breuer}}, \bibinfo {author} {\bibfnamefont {E.-M.}\ \bibnamefont {Laine}}, \
  and\ \bibinfo {author} {\bibfnamefont {J.}~\bibnamefont {Piilo}},\ }\href
  {\doibase 10.1103/PhysRevLett.103.210401} {\bibfield  {journal} {\bibinfo
  {journal} {Phys. Rev. Lett.}\ }\textbf {\bibinfo {volume} {103}},\ \bibinfo
  {pages} {210401} (\bibinfo {year} {2009})}\BibitemShut {NoStop}%
\bibitem [{\citenamefont {Glaser}\ \emph {et~al.}(2015)\citenamefont {Glaser},
  \citenamefont {Boscain}, \citenamefont {Calarco}, \citenamefont {Koch},
  \citenamefont {K{\"o}ckenberger}, \citenamefont {Kosloff}, \citenamefont
  {Kuprov}, \citenamefont {Luy}, \citenamefont {Schirmer}, \citenamefont
  {Schulte-Herbr{\"u}ggen}, \citenamefont {Sugny},\ and\ \citenamefont
  {Wilhelm}}]{Glaser2015}%
  \BibitemOpen
  \bibfield  {author} {\bibinfo {author} {\bibfnamefont {S.~J.}\ \bibnamefont
  {Glaser}}, \bibinfo {author} {\bibfnamefont {U.}~\bibnamefont {Boscain}},
  \bibinfo {author} {\bibfnamefont {T.}~\bibnamefont {Calarco}}, \bibinfo
  {author} {\bibfnamefont {C.~P.}\ \bibnamefont {Koch}}, \bibinfo {author}
  {\bibfnamefont {W.}~\bibnamefont {K{\"o}ckenberger}}, \bibinfo {author}
  {\bibfnamefont {R.}~\bibnamefont {Kosloff}}, \bibinfo {author} {\bibfnamefont
  {I.}~\bibnamefont {Kuprov}}, \bibinfo {author} {\bibfnamefont
  {B.}~\bibnamefont {Luy}}, \bibinfo {author} {\bibfnamefont {S.}~\bibnamefont
  {Schirmer}}, \bibinfo {author} {\bibfnamefont {T.}~\bibnamefont
  {Schulte-Herbr{\"u}ggen}}, \bibinfo {author} {\bibfnamefont {D.}~\bibnamefont
  {Sugny}}, \ and\ \bibinfo {author} {\bibfnamefont {F.~K.}\ \bibnamefont
  {Wilhelm}},\ }\href {\doibase 10.1140/epjd/e2015-60464-1} {\bibfield
  {journal} {\bibinfo  {journal} {Eur. Phys. J. D}\ }\textbf {\bibinfo {volume}
  {69}},\ \bibinfo {pages} {279} (\bibinfo {year} {2015})}\BibitemShut
  {NoStop}%
\bibitem [{\citenamefont {Riahi}\ \emph {et~al.}(2016)\citenamefont {Riahi},
  \citenamefont {Salomon}, \citenamefont {Glaser},\ and\ \citenamefont
  {Sugny}}]{PhysRevA.93.043410}%
  \BibitemOpen
  \bibfield  {author} {\bibinfo {author} {\bibfnamefont {M.~K.}\ \bibnamefont
  {Riahi}}, \bibinfo {author} {\bibfnamefont {J.}~\bibnamefont {Salomon}},
  \bibinfo {author} {\bibfnamefont {S.~J.}\ \bibnamefont {Glaser}}, \ and\
  \bibinfo {author} {\bibfnamefont {D.}~\bibnamefont {Sugny}},\ }\href
  {\doibase 10.1103/PhysRevA.93.043410} {\bibfield  {journal} {\bibinfo
  {journal} {Phys. Rev. A}\ }\textbf {\bibinfo {volume} {93}},\ \bibinfo
  {pages} {043410} (\bibinfo {year} {2016})}\BibitemShut {NoStop}%
\bibitem [{\citenamefont {Shu}\ \emph {et~al.}(2016)\citenamefont {Shu},
  \citenamefont {Ho},\ and\ \citenamefont {Rabitz}}]{PhysRevA.93.053418}%
  \BibitemOpen
  \bibfield  {author} {\bibinfo {author} {\bibfnamefont {C.-C.}\ \bibnamefont
  {Shu}}, \bibinfo {author} {\bibfnamefont {T.-S.}\ \bibnamefont {Ho}}, \ and\
  \bibinfo {author} {\bibfnamefont {H.}~\bibnamefont {Rabitz}},\ }\href
  {\doibase 10.1103/PhysRevA.93.053418} {\bibfield  {journal} {\bibinfo
  {journal} {Phys. Rev. A}\ }\textbf {\bibinfo {volume} {93}},\ \bibinfo
  {pages} {053418} (\bibinfo {year} {2016})}\BibitemShut {NoStop}%
\bibitem [{\citenamefont {Guo}\ \emph {et~al.}(2019)\citenamefont {Guo},
  \citenamefont {Luo}, \citenamefont {Ma},\ and\ \citenamefont
  {Shu}}]{PhysRevA.100.023409}%
  \BibitemOpen
  \bibfield  {author} {\bibinfo {author} {\bibfnamefont {Y.}~\bibnamefont
  {Guo}}, \bibinfo {author} {\bibfnamefont {X.}~\bibnamefont {Luo}}, \bibinfo
  {author} {\bibfnamefont {S.}~\bibnamefont {Ma}}, \ and\ \bibinfo {author}
  {\bibfnamefont {C.-C.}\ \bibnamefont {Shu}},\ }\href {\doibase
  10.1103/PhysRevA.100.023409} {\bibfield  {journal} {\bibinfo  {journal}
  {Phys. Rev. A}\ }\textbf {\bibinfo {volume} {100}},\ \bibinfo {pages}
  {023409} (\bibinfo {year} {2019})}\BibitemShut {NoStop}%
\bibitem [{\citenamefont {Hong}\ \emph {et~al.}(2021)\citenamefont {Hong},
  \citenamefont {Fan}, \citenamefont {Shu},\ and\ \citenamefont
  {Henriksen}}]{PhysRevA.104.013108}%
  \BibitemOpen
  \bibfield  {author} {\bibinfo {author} {\bibfnamefont {Q.-Q.}\ \bibnamefont
  {Hong}}, \bibinfo {author} {\bibfnamefont {L.-B.}\ \bibnamefont {Fan}},
  \bibinfo {author} {\bibfnamefont {C.-C.}\ \bibnamefont {Shu}}, \ and\
  \bibinfo {author} {\bibfnamefont {N.~E.}\ \bibnamefont {Henriksen}},\ }\href
  {\doibase 10.1103/PhysRevA.104.013108} {\bibfield  {journal} {\bibinfo
  {journal} {Phys. Rev. A}\ }\textbf {\bibinfo {volume} {104}},\ \bibinfo
  {pages} {013108} (\bibinfo {year} {2021})}\BibitemShut {NoStop}%
\bibitem [{\citenamefont {Fan}\ \emph {et~al.}(2023)\citenamefont {Fan},
  \citenamefont {Shu}, \citenamefont {Dong}, \citenamefont {He}, \citenamefont
  {Henriksen},\ and\ \citenamefont {Nori}}]{PhysRevLett.130.043604}%
  \BibitemOpen
  \bibfield  {author} {\bibinfo {author} {\bibfnamefont {L.-B.}\ \bibnamefont
  {Fan}}, \bibinfo {author} {\bibfnamefont {C.-C.}\ \bibnamefont {Shu}},
  \bibinfo {author} {\bibfnamefont {D.}~\bibnamefont {Dong}}, \bibinfo {author}
  {\bibfnamefont {J.}~\bibnamefont {He}}, \bibinfo {author} {\bibfnamefont
  {N.~E.}\ \bibnamefont {Henriksen}}, \ and\ \bibinfo {author} {\bibfnamefont
  {F.}~\bibnamefont {Nori}},\ }\href {\doibase 10.1103/PhysRevLett.130.043604}
  {\bibfield  {journal} {\bibinfo  {journal} {Phys. Rev. Lett.}\ }\textbf
  {\bibinfo {volume} {130}},\ \bibinfo {pages} {043604} (\bibinfo {year}
  {2023})}\BibitemShut {NoStop}%
\end{thebibliography}%
\end{document}